\documentclass[
10pt,
preprint,
superscriptaddress,
amsmath,amssymb,
aps,
prfluids
]{revtex4-2}

\usepackage{graphicx}
\usepackage{dcolumn}
\usepackage{bm}
\usepackage{placeins}
\usepackage[dvipsnames]{xcolor}
\usepackage{soul}
\usepackage{caption}
\usepackage{subcaption}
\usepackage{comment}
\usepackage{booktabs} 

\usepackage{multirow}
\usepackage{soul}

\begin{document}

 \title[Article Title]{
Nonhomogeneous elastic turbulence in the two-dimensional Taylor-Couette flow
}

\author{Zhongxuan Hou}
 \thanks{Corresponding author}\email{zhongxuan.hou@ensam.eu}
\affiliation{Univ. Lille, CNRS,
ONERA, Arts et M\'etiers Institute of Technology, Centrale Lille, UMR 9014-LMFL-Laboratoire de Mécanique des Fluides de
Lille - Kamp\'e de F\'eriet, F-59000, Lille, France}

 \author{Stefano Berti}
  \affiliation{Univ. Lille, ULR 7512 - Unit\'e de M\'ecanique de Lille Joseph Boussinesq (UML), F-59000 Lille, France}

 \author{Teodor Burghelea}
  \affiliation{Laboratoire de Thermique et Energie de Nantes, CNRS, Nantes Universit\'e, Rue Christian Pauc, B.P. 50609, Nantes, 44306, France}

  \author{Francesco Roman\`o}
  \affiliation{Univ. Lille, CNRS,
ONERA, Arts et M\'etiers Institute of Technology, Centrale Lille, UMR 9014-LMFL-Laboratoire de Mécanique des Fluides de
Lille - Kamp\'e de F\'eriet, F-59000, Lille, France}
 
\begin{abstract}
Elastic turbulence is a spatially and temporally disordered flow state appearing in viscoelastic fluids at vanishing fluid inertia and large elasticity. The resulting flows have broad technological interest, particularly to enhance mixing and heat transfer in microdevices. Although its experimental characterization is now well established in different setups, its theoretical understanding and numerical reproducibility remain challenging, especially in wall-bounded geometries. By means of numerical simulations, we investigate the onset of elastic turbulence and the characteristics of the developed turbulent-like states in the two-dimensional, confined, Taylor-Couette system. First, we characterize the purely elastic instability, addressing previously contrasting evidence. We then show that the fully nonlinear dynamics are weakly anisotropic and strongly nonhomogeneous. Indeed, they are confined in a dynamically active region adjacent to the inner wall, akin to the elastic boundary layer from previous predictions. Within this region, in spite of some non-negligible deviations due to the nonhomogeneity of our setup, the statistical and spectral turbulent properties are 
to a reasonable extent not far from the theoretical expectations and experimental observations.
\end{abstract}
\maketitle

\section{Introduction}\label{sec:intro}
Viscoelastic fluids, such as dilute polymer solutions, generate elastic stresses, associated with polymer stretching, which add to the viscous ones present also in Newtonian fluids. At small Reynolds numbers ($Re$), where nonlinear inertial effects are negligible, elastic stresses are the primary source of flow destabilization. Via their feedback effect on the carrying fluid, indeed these extra stresses can give rise to purely elastic instabilities, often controlled by normal stress differences~\cite{larson_1990,larson_instabilities_1992,groisman_2000,steinberg2021}. When the Weissenberg number ($Wi$)~-~the product of a characteristic polymer relaxation time and a typical flow shear rate~-~exceeds a critical value, meaning for large enough elasticity, such instabilities can develop and drive the flow into a temporally and spatially disordered state known as elastic turbulence~\cite{groisman_2000,groisman_2004,steinberg2021}. The latter dynamics display some resemblance with Newtonian, inertial turbulence, including efficient transport properties. Therefore, beyond its fundamental relevance, elastic turbulence also appears interesting for applications, particularly in microfluidic devices. Indeed, it has been shown to efficiently enhance mixing~\cite{groisman2001efficient} and heat transfer~\cite{traore2015efficient,abed2016experimental} in flows at low Reynolds number.

Since its discovery in the von Kármán swirling flow between two disks~\cite{groisman_2000}, elastic turbulence and the instabilities at its origin have been thoroughly characterized experimentally in a variety of setups, such as the Taylor-Couette flow~\cite{groisman_2004,zhang2025experimental}, the serpentine channel flow~\cite{groisman2001efficient,Duclou_2019}, and the cross-slot flow~\cite{arratia2006elastic,sousa2015purely}. Most of these systems are characterized by curvilinear streamlines, where the generation of secondary radial flow is understood in terms of the hoop stress mechanism~\cite{groisman_1998,schaefer2018,steinberg2021}. In a nutshell, the coupling between normal stress differences and the curvature of the base flow results in a net force in the radial direction. The curvature dependence of the purely elastic instability was investigated in the three-dimensional (3D) Taylor-Couette system in a recent study~\cite{schaefer2018} experimentally and analytically, showing that the onset is captured by an appropriate extension of the Pakdel-McKinley criterion to finite gap widths. Note that purely elastic instabilities and ensuing turbulent-like behavior have been documented also in parallel flows characterized by straight streamlines, in theoretical~\cite{morozov2007introductory,morozov2019subcritical}, numerical~\cite{berti_two-dimensional_2008,berti_elastic_2010, plan2017lyapunov,gupta2019effect,lellep2024purely} and experimental work~\cite{pan2013nonlinear}. However, their detailed discussion goes beyond the scope of the present study.

In various flow configurations such as von Karman swirling flow, Taylor-Couette flow, curvilinear channels, and common rheometric flows, the main features of elastic turbulence are the increase of both flow resistance and mixing efficiency beyond a critical Weissenberg number, and the concomitant appearance of a complex, multiscale flow structure, as revealed by power-law energy spectra reminiscent of Newtonian turbulence. More recently it has been found that features characteristic of elastic turbulence can be observed also in inertial turbulent flows of viscoelastic fluids at small scales~\cite{Garg2025}. In the majority of experiments (see~\cite{steinberg2021} for a review) and numerical simulations (see, e.g.,~\cite{berti_two-dimensional_2008,berti_elastic_2010,Grilli_2013}), however, kinetic energy spectra of the bulk flow, in both the frequency and wavenumber domains, are found to be steeper than in 3D inertial turbulence, with a power-law exponent larger than $3$, in absolute value. This points to a temporally disordered flow mainly controlled by the largest length scales. The experimental and numerical measures match the theoretical prediction obtained in homogeneous isotropic conditions~\cite{steinberg2021}, according to which the wavenumber kinetic energy spectrum scales as $\mathcal{E}_k(k) \sim k^{-\sigma_k}$, with $\sigma_k>3$~\cite{fouxon_spectra_2003}. Note, however, that most often in experiments wavenumber spectra are obtained from frequency ones, $\mathcal{E}_k(f)$, by application of Taylor's hypothesis. Previous work documented that this hypothesis may be only partially valid in elastic turbulence~\cite{Burghelea_2005,garg2021statistical}. The same theory also provides the scaling of the elastic energy spectrum $\mathcal{E}_e(k) \sim k^{-\sigma_e}$, with $\sigma_e=\sigma_k-2$~\cite{fouxon_spectra_2003,steinberg2021}. It is worth noting, however, that other, recent, theoretical arguments suggest that $\sigma_e=(\sigma_k-1)/2$~\cite{singh2024intermittency}. While measuring elastic energy spectra is difficult in experiments, this can be done in numerical simulations. From the latter, depending on the configuration, different values of $\sigma_e$ have been reported~\cite{nguyen2016small,garg2018particle,lellep2024purely,singh2024intermittency,beneitez2024transition}, but no full agreement with either prediction has emerged. The determination of the scaling of $\mathcal{E}_e(k)$ then still appears an open question.

In confined geometry, the structure and the statistical properties of the flow should be modified by the presence of boundaries. Experiments indicate that within the boundary layer velocity gradients are much more important than in the bulk, suggesting that this leads to an accumulation of elastic stresses there, which later propagate further away from the wall~\cite{Burghelea_2007,Jun_2011}. To our knowledge, the nonhomogeneity of elastic turbulence induced by the presence of walls is still only poorly explored theoretically and numerically.

In this work we numerically investigate the two-dimensional (2D) Taylor-Couette flow in the limit of vanishing $Re$ and for progressively larger fluid elasticity. While previous studies already examined purely elastic instabilities and the high-$Wi$ regime in this system~\cite{buel_elastic_2018,van2020active}, some results did not appear completely conclusive. In particular, some evidence supported the picture of a supercritical instability~\cite{buel_elastic_2018}, while some others pointed to a subcritical one~\cite{vanBuel_2024}. Moreover, while the exponents of kinetic energy spectra were measured as a function of the radial coordinate in the gap between the two cylinders~\cite{buel_elastic_2018}, the spatial nonhomogeneity of the flow remains to be explored in detail. This seems to us an important point, both at a fundamental level and in view of the characterization of its effects on mixing properties, and is thus the main focus of our study.  

The choice of a 2D setup is here intended as a first step in these directions. Furthermore, to some extent, it is supported by the lack of observations of significant differences between the main statistical features of elastic turbulence in 2D numerical simulations and 3D experiments (see, e.g.,~\cite{berti_two-dimensional_2008, gupta2019effect,yerasi2024preserving}). In the same spirit, we consider the simplest possible viscoelastic fluid model, corresponding to linear elasticity of polymer chains. In our numerical work, particular attention was paid to ensuring the robustness of results with respect to the spatial and temporal resolution of the simulations. An effort is also made to compare the present results with those from 3D numerical simulations~\cite{Song_2022,song2023self}, and experiments~\cite{groisman_2004,zhang2025experimental}. This will allow us to document the similarities and differences of our simplified 2D setup with realistic geometries. If space dimensionality likely has a non-negligible role in the instability pathway, its effects on the phenomenology of turbulent-like states might be expected to be less important, with possible quantitative differences but similar qualitative features. In this regard, we note that the Taylor-Couette flow has some peculiarities. Early experiments~\cite{groisman_2004} indicate that the kinetic energy spectrum (at mid gap) may be flatter than in other geometries and with respect to the theoretical expectation. Moreover, measurements (in the frequency domain) show that it is characterized by two distinct power-law decay regions (with exponents $\zeta_k= 1.1$ and $2.2$ at low and high frequencies, respectively). The inflection frequency is found to be not far from the rotation rate driving the fluid in motion. In addition, it is argued that the observed spectral behavior may be the consequence of axisymmetric coherent structures that dominate the low-frequency part of the spectrum. Studies based on 3D numerical simulations at high $Wi$ (but not always negligible $Re$), also report on inflection frequencies~\cite{Thomas_2009} or wavenumbers~\cite{song2023self}, but find more varied spectral slopes~\cite{Thomas_2009,Liu_Khomami_2013,Song_2022,song2023self}, in the frequency and wavenumber domains, not always necessarily in quantitative agreement with the experimental ones. Finally, it must be noted that other, recent experiments instead report about temporal spectra at mid gap characterized by a single exponent $\zeta_k \approx 3$~\cite{zhang2025experimental}.

This article is organized as follows. In Sec.~\ref{sec:problem_formulation}, the mathematical formulation of the problem is introduced. There, we separately present the equations governing the dynamics of the viscoelastic fluid (Sec.~\ref{sec:gov_eqs}), and the numerical method and settings used for our simulations (Sec.~\ref{sec:numerical_method}). The results are illustrated in Sec.~\ref{sec:results}. We start from the characterization of the purely elastic instability (Sec.~\ref{sec:instability}), and of global properties of the developed turbulent-like states (Sec.~\ref{sec:global}). We then quantify the spatial nonhomogeneity of the elastic-turbulence flows in Sec.~\ref{sec:boundary_layer}, and analyze their energy spectra in Sec.~\ref{sec:spectra}. Conclusions are presented in Sec.~\ref{sec:concl}.

\section{Problem formulation}\label{sec:problem_formulation}

\subsection{Governing equations}\label{sec:gov_eqs}
\begin{figure}[!t]
\centering
\subfloat[][Viscoelastic Taylor--Couette]{\includegraphics[height=0.25\textwidth]{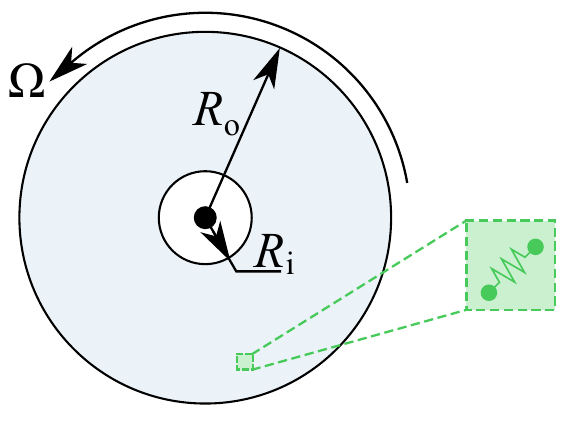}}\qquad \qquad  
\subfloat[][Computational mesh]{\qquad \includegraphics[height=0.25\textwidth]{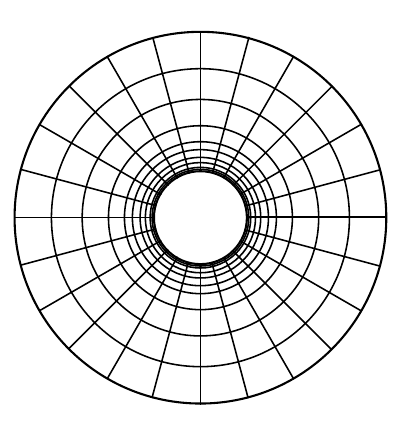}\qquad }
\caption{(a) Schematic of the Taylor-Couette viscoelastic setup. The inner wall is fixed, while the outer wall rotates in counter-clockwise direction at peripheral velocity $\Omega R_o$. The radial gap parameter is $\eta = R_i/R_o = 1/4$. (b) Schematic of a typical distribution for the numerical mesh: cells are uniformly distributed in the azimuthal direction and expand in the radial one by an overall factor of approximately $9.5$.}
\label{fig:Schematics}
\end{figure}

The two-dimensional Taylor-Couette setup is considered to investigate the viscoelastic flow under creeping flow conditions. The flow is driven by the outer cylinder of radius $R_o$ rotating at peripheral velocity $\Omega R_o$. The inner cylinder of radius $R_i = R_o/4$ is kept at rest. A schematic of the corresponding Taylor-Couette setup is depicted in Fig.~\ref{fig:Schematics}a.

The viscoelastic fluid is assumed to be incompressible and is characterized by the density $\rho$, the solvent viscosity $\mu_s$ and the polymeric viscosity $\mu_p$, whose sum results in the total fluid viscosity $\mu = \mu_s + \mu_p$. Consistently, the total stress tensor $\bm{\tau}$ is due to a solvent $\bm{\tau}_s$ and a polymeric $\bm{\tau}_p$ contribution, i.e. $\bm{\tau} = \bm{\tau}_s + \bm{\tau}_p$. The solvent is assumed Newtonian, hence $\bm{\tau}_s = \mu_s\left[\bm{\nabla} \bm{u} + (\bm{\nabla} \bm{u})^T\right]$, where $\bm{u}$ denotes the velocity field. The Oldroyd-B model~\cite{oldroyd_formulation_1950} is employed to characterize the extra stresses resulting from the viscoelastic behavior of dilute polymer solutions. Two major limitations are associated with this model. The first one is that the polymeric response is assumed to be characterized by a single relaxation time $\lambda$. The second limitation comes from the assumption of linear elasticity, namely the choice to model polymers as infinitely extensible Hookean massless dumbbells~\cite{larson_instabilities_1992}. While both such limitations can be overcome by more complex rheological models, here we adopt the Oldroyd-B model as it provides the simplest possible description (including the upper-convected Maxwell model in the limit $\mu_p \gg \mu_s$) accounting for elastic instabilities and, ultimately, elastic turbulence~\cite{berti_two-dimensional_2008,Shaqfeh_2021,Sanchez_2022,Burghelea_2019} 

The dynamics of the system are governed by the continuity and momentum-conservation equations, supplemented by the Oldroyd-B constitutive law. Choosing $R_o$, $1/\Omega$, $\Omega R_o$, $\mu_p\Omega$, and $\mu \Omega$, as typical length, time, velocity, polymeric stress, and pressure scales, respectively, the equations of motion, in nondimensional form, read: 
\begin{subequations}
\begin{align}
    \nabla\cdot\bm{u} &= 0, \label{eqn:cont}\\
    Re\bigg(\frac{\partial  \bm{u}}{\partial t} +  \bm{u}\cdot \bm{\nabla}  \bm{u}\bigg) &= - \bm{\nabla} \bm{p} \ +\ \beta \bm{\nabla}^2 \bm{u} + (1-\beta)\bm{\nabla} \cdot \bm{\tau_p}, \label{eqn:mom_BSD}\\
    \bm{\tau}_p + Wi\bigg[\frac{\partial \bm{\tau}_p}{\partial t} + \bm{u} \cdot \bm{\nabla} \bm{\tau}_p -(\bm{\nabla}  
        \bm{u})^T \bm{\tau}_p - \bm{\tau}_p \bm{\nabla} \bm{u}\bigg] &= \bm{\nabla} \bm{u} + \bm{\nabla}^T \bm{u},
     \label{eqn:const}
\end{align}    
\end{subequations}
where $t$ denotes time, $p$ pressure, and $\bm{u} = (u_r,\ u_\phi)$ is the velocity field (the subscripts $r$ and $\phi$ indicating the radial and azimuthal components). Besides the radial gap $\eta=R_i/R_o$, three nondimensional parameters control the dynamics of the problem, namely the Reynolds number $Re = \rho R_o^2\Omega/\mu$, the solvent-to-total viscosity ratio $\beta = \mu_s/\mu$, and the Weissenberg number $Wi=\lambda \Omega$. 
Note that, alongside the Weissenberg number, fluid-elasticity effects are also accounted for by the Deborah number, $De = \lambda / \tau_f$, where $\tau_f$ is a typical flow timescale, which here can be taken as $\tau_f=\Omega^{-1}$. Therefore, within the present configuration, these two nondimensional parameters are equivalent ($Wi \equiv De$).


In this study, we will focus on the effect of the Weissenberg number, varying it in the range $0 \leq Wi \leq 100$, and keeping fixed the other nondimensional groups to the values $Re = 2.5\times 10^{-4}$, $\beta = 2/5$, $\eta=1/4$. We stress that these definitions and ranges are compatible with those used in previous studies (see, e.g.,~\cite{buel_elastic_2018}), which will be referred to in the following. The mathematical formulation is completed by no-slip velocity boundary conditions at the inner and outer walls, $\bm{u}(r=\eta) = \bm{0}$ and $\bm{u}(r=1) = \bm{e}_\phi$ (with $\bm{e}_\phi$  the unit vector in the azimuthal direction). A linear extrapolation technique is adopted for the extra-stress tensor $\bm{\tau}_p$~\cite{pimenta_stabilization_2017} at $r=1$, while homogeneous Neumann boundary conditions are set at $r=\eta$. As for the initial condition, unless otherwise stated, the results presented in Sec.~\ref{sec:results} are obtained starting from a no-flow state for both the velocity and the stress.
\subsection{Numerical method}\label{sec:numerical_method}
We use the open-source code  OpenFOAM\textsuperscript \textregistered \ with the RheoTool\textsuperscript \textregistered \ viscoelastic-solver package to perform finite-volume numerical time-marching simulations of Eqs.~(\ref{eqn:cont})-(\ref{eqn:const}). To control numerical instabilities associated with large stress gradients, appearing at large $Wi$, the log-conformation technique~\cite{Fattal_2004,alves_numerical_2021} is adopted. This amounts to replacing the integration of Eq.~(\ref{eqn:const}) with that of the evolution equation of the field $\bm{\Theta} = \ln(\bm{C}) = \bm{R}\ln(\bm{A})\bm{R}^T$. In the last expression, $\bm{A}$ denotes the diagonal form of the (positive-definite) conformation tensor $\bm{C}$ and $\bm{R}$ is the basis-change (orthogonal)  matrix needed to diagonalize $\bm{C}$. The conformation tensor is related to the polymeric stress by $\bm{\tau}_p =Wi^{-1} \, (\bm{C}-\bm{I})$, where $\bm{I}$ is the identity matrix. This approach allows one to remedy the loss (due to numerical instabilities at high $Wi$) of the positive-definiteness constraint for $\bm{C}$~\cite{favero_viscoelastic_2010}. The other numerical details employed in RheoTool, such as linear solvers and preconditioners, are reported in Table~\ref{tab:OpenFOAM}. Finally, note that all the results presented in the following have been computed in double precision.

The sensitivity of the results presented below to the choices of the solvers and preconditioners was carefully checked. We observed that the onset and dynamics of destabilizing flow patterns are significantly affected by such numerical details if the spatial resolution is not sufficient to grant grid convergence. A dedicated grid-convergence study on the most relevant observables considered in this work is reported in Appendix~A. Four meshes have been tested evenly distributing the finite volumes in the azimuthal direction and using a constant cell expansion ratio of 9.5 in the radial direction (see Fig.~\ref{fig:Schematics}b). Letting $N_r$ and $N_\phi$ be the number of cells in the radial and azimuthal direction, respectively, Appendix~A compares the results for $N_r \times N_{\phi} = 100 \times 120$ (same as in~\cite{buel_elastic_2018,vanBuel_2024}), $150 \times 240$, $200 \times 360$, and $250 \times 480$. The convergence with respect to the temporal resolution was also tested, by considering the time steps $\Delta t = (1.3, 3.1, 6.3) \times 10^{-5}$ (see Appendix~A).
Based on those analyses, the combination of spatial and temporal resolutions given by $N_r \times N_{\phi} = 200 \times 360$ and $\Delta t = 6.3\times 10^{-5}$ ensures grid and time step independent solutions of the flow physics and statistics discussed in this article (see Appendix~A). Moreover, to make sure that the statistical characterizations of the flow in the elastic-turbulence regime are well converged, we always simulate the dynamics for at least $40$ polymer relaxation times in the statistically steady state, adapting therefore the simulation duration according to the value of $Wi$. 

\begin{table}[!h]
\begin{ruledtabular}
\begin{tabular}{lcc}
\textrm{Equations}&
\textrm{Solvers}&
\textrm{Preconditioner}\\
\colrule
Momentum & PCG solver & Incomplete Cholesky  \\
    Continuity & PCG solver & Incomplete Cholesky  \\
    Oldroyd-B & PBiCG &Diagonal Incomplete-LU \\
\end{tabular}
\caption{Details of the solver and preconditioner options selected in the simulations.}
    \label{tab:OpenFOAM}
\end{ruledtabular}
\end{table}

\section{Results}\label{sec:results}
In this section we report the results of our numerical simulations. We will first focus on the characterization of the purely elastic instability and then discuss the turbulent-like features for flows at Weissenberg numbers larger than the critical one. Specifically, we will examine their isotropy and homogeneity properties, highlighting the existence of a boundary-layer region where the nonlinear dynamics are mostly concentrated, and then turn to the analysis of energy spectra.

\subsection{Basic state and flow instability} \label{sec:instability}

The viscoelastic Taylor--Couette flow, described by the Oldroyd-B model in Eqs.(\ref{eqn:mom_BSD})-(\ref{eqn:const}), admits a steady basic state that can be computed analytically. The velocity field corresponds to the one for the Newtonian Taylor--Couette system, while the extra stresses are self-balanced up until an instability sets in. Following Ref.~\cite{larson_1990}, the basic-state velocity $\bm{u}^0$ and extra stresses $\bm{\tau}^0_{p}$ read:
\begin{subequations}
\begin{align}
    u_r^0 &= 0,\quad u_\phi^0 = \cfrac{r}{1-\eta^2} - \cfrac{\eta^2}{r\left(1-\eta^2\right)} \label{eqn:basic_vel}\\
    \tau^0_{p,rr} &= 0, \quad \tau^0_{p,r\phi} = \cfrac{2\eta^2}{r^2\left(1-\eta^2\right)}, \quad \tau^0_{p,\phi\phi} =  \cfrac{8\,Wi\,\eta^4}{r^4\left(1-\eta^2\right)^2}. \label{eqn:basic_stress}
\end{align}
\end{subequations}
Note that such a steady solution is independent of $\phi$. Moreover, the elastic stress vanishes either with the second ($\tau^0_{p,r\phi}$) or the fourth ($\tau^0_{p,\phi\phi}$) power of $r$, and its component $\tau^0_{p,\phi\phi}$ is directly proportional to $Wi$. 

\begin{figure}[p]
\vspace{-0.5cm}
\centering
\includegraphics[width=0.675\linewidth]{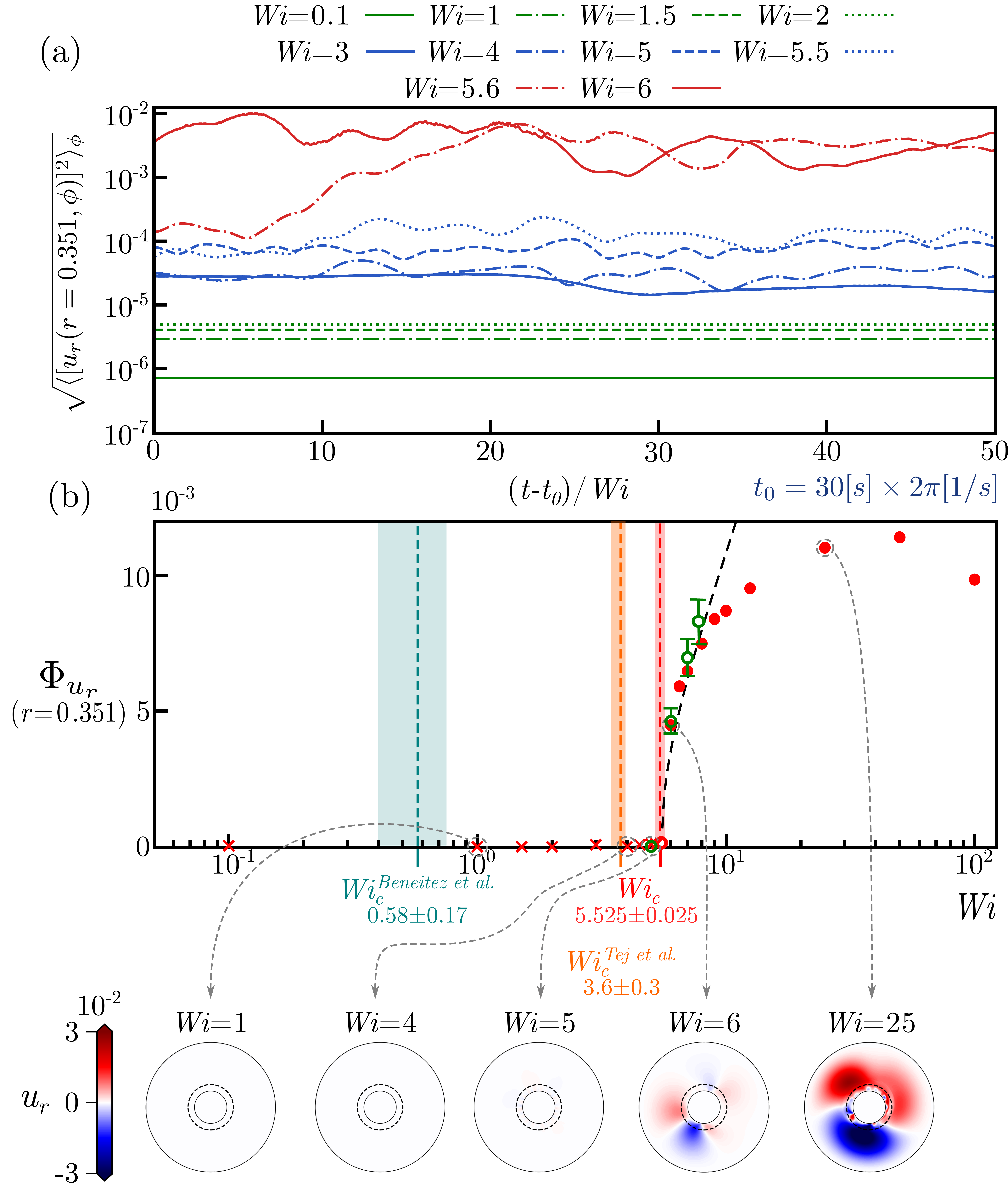} \vspace{-0.5cm}
\caption{(a) Absolute value of the radial velocity $u_r$ evaluated at $r=0.351$ and averaged over $\phi$ for various subcritical and weakly supercritical Weissenberg numbers. (b) The order parameter $\Phi_{u_r}$ evaluated at $r=0.351$ for different $Wi$ by ramping up in $Wi$ starting from rest (red markers) or ramping down in $Wi$ starting from fully-developed conditions at $Wi = 50$ (green markers, with a tolerance range of 10\%). The red crosses ($Wi \leq 5$) denote stable conditions, the empty red circle is within the tolerance range across the critical Weissenberg number $Wi_c = 5.525\pm 0.025$, while the full red markers denote unstable conditions. The black dashed line corresponds to a best fit with $(Wi-Wi_c)^{1/2}$. The visualizations of the radial velocity field at the bottom demonstrate the significance of the secondary flow over the domain. The dashed line in these $u_r$ snapshots represents the line on which the order parameter $\Phi_{u_r}$ is computed (in this case $r=0.351$). The decrease of $\Phi_{u_r}(r=0.351)$ for $Wi>25$ is due to the outward radial shift of the secondary flow extrema (compare $Wi=6$ and $Wi=25$). The critical Weissenberg number predicted by \cite{Beneitez2025} ($Wi^{Beneitez\ et\ al.}_c$) and \cite{Tej2025} ($Wi^{Tej\ et\ al.}_c$) for PDI is depicted in light-blue and orange color, respectively, after adjusting their definitions of $Wi$ to our scaling.}
\label{fig:Secondary_flow_strength}
\end{figure}

Upon an increase of the Weissenberg number, the flow undergoes an instability~\cite{larson_1990,larson_instabilities_1992,buel_elastic_2018} that leads to radial elongational flow, as well as to non-zero $\tau_{p,rr}$. A first indication on the accompanying intensity of the secondary flow can be appreciated from  Fig.~\ref{fig:Secondary_flow_strength}a, which
shows the time evolution of the absolute value of the azimuthally averaged radial velocity $\sqrt{\langle[u_r(r,\phi,t)]^2\rangle_{\phi}}$ evaluated at $r=0.351$ for various subcritical and weakly supercritical Weissenberg numbers after $t=188$ (in nondimensional units). Here, $\langle\cdot\rangle_\phi$ denotes an azimuthal average, and the choice of computing $\Phi_{u_r}$ at $r=0.351$ is made to optimize the receptivity of the secondary flow for the lowest supercritical Weissenberg numbers, which tend to induce the largest intensity of the secondary flow in the interval $r\in [0.33,\ 0.45]$. 
We stress that selecting other radial coordinates (not shown) leads to the same conclusions drawn in the following. At $Wi<3$, the numerics indicate that the secondary flow remains essentially stationary
and at a very low intensity (amplitudes smaller than $10^{-5}$). For such Weissenberg numbers, the flow presents numerically negligible deviations from the purely azimuthal base flow. As $Wi$ increases, i.e. for $3<Wi<5.5$, some finite-amplitude temporal fluctuations of $u_r$ emerge, indicating the progressive development of a small radial velocity from the laminar state. Owing to the low intensity of such radial flow, we cannot numerically assess whether the origin of such a small, yet finite, radial perturbation is due to numerical errors or other sorts of numerically-induced instabilities, such as the PDI reported by \cite{Beneitez2025,Tej2025}. While the secondary flow remains always around $10^{-4}$ in amplitude for $Wi \leq 5.5$, a clear transition occurs for $Wi = 5.6$, where the secondary flow increases by two orders of magnitude in amplitude, passing from $10^{-4}$ to $10^{-2}$ (see Fig.~\ref{fig:Secondary_flow_strength}a). The qualitative difference between $Wi=5.5$ and $Wi=5.6$ demonstrates that these two Weissenberg numbers bracket the critical one for a flow instability onset. At $Wi = 6$, these fluctuations become sustained, indicating the establishment of a significant secondary flow in the radial direction. This progressive change suggests that our system carries two-dimensional perturbations of numerical origin that are small [at most of $\mathcal{O}(10^{-4})$] but cannot be regarded as infinitesimal, which then get amplified by two orders of magnitude upon the instability onset.

To investigate the onset and characteristics of such instability, we employ a slightly different order parameter $\Phi_{u_r}$ with respect to the previously defined indicator, namely:
\begin{equation}\label{eqn:secondary_flow_strength}
    \Phi_{u_r}(r)=\sqrt{\overline{\langle[u_r(r,\phi,t)]^2\rangle_{\phi}}},
\end{equation}  
where $\overline{(\cdot)}$ denotes 
a temporal average. To avoid interpolation errors, the order parameter is evaluated at the velocity collocation points. We start measuring $\Phi_{u_r}$, at $r=0.351$, after $t=600$ to make sure that the transient from the initial conditions has become negligible. The time average used to compute $\Phi_{u_r}$ is taken over a time span longer than $40 Wi$ in nondimensional time units (corresponding to 40 relaxation times $\lambda$ of the polymer). 

Figure~\ref{fig:Secondary_flow_strength}b then shows this order parameter for $0.1 \leq Wi \leq 100$. The red markers denote the time-converged results from simulations starting from rest. The black dashed line corresponds to a square-root best fit of $\Phi_{u_r}(r=0.351)$ against $Wi$ for weakly supercritical conditions, namely for $ 0.01 \leq Wi-Wi_c \leq 2.5$. The asymptotic fit is theoretically consistent with linear instabilities leading to forward bifurcation, i.e. $\Phi_{u_r}\sim \sqrt{Wi-Wi_c}$, and in good agreement with our data (compare the red markers and the black dashed line in Fig.~\ref{fig:Secondary_flow_strength}). Note, however, that the linear nature of this instability remains to be proven. By power-law regression (details in the caption of Fig.~\ref{fig:Secondary_flow_strength}), we could identify the critical Weissenberg number $Wi_c = 5.525\pm0.025$ as the value of the control parameter such that the order parameter $\Phi_{u_r} $becomes smaller than the accuracy we can guarantee, i.e. $10^{-4}$.

To confirm the above critical threshold (in $Wi$), we also tested slightly supercritical Weissenberg numbers, i.e. $Wi = 5.55$, $5.6$, $5.75$, which are found to be unstable. Due to their proximity to $Wi_c$, we did not include the corresponding data in the regression, because in those cases much longer time series would be needed to converge to a fully developed state. This is typical for linear instabilities, as it is due to the so-called critical slowing down, a phenomenon that implies exponentially diverging relaxation times for small perturbations if $Wi \rightarrow Wi_c$. Finally, we further provide a visual confirmation that our computation of $Wi_c$ based on $\Phi_{u_r}(r=0.351)$ is reliable. Specifically, in Fig.~\ref{fig:Secondary_flow_strength}, we also show the radial velocity field for $Wi = 5$, $6$ and $25$ all over the domain. These three snapshots provide a qualitative understanding of the robustness of our criterion. In fact, one can readily remark that no significant perturbation is observed for $Wi = 5 \approx 0.9\times Wi_c$, unlike for $Wi = 6 \approx 1.1\times Wi_c$, where the secondary flow strength in the radial direction is $\approx 5\times 10^{-3}$. 

To further explore the type of bifurcation identified, we carried out four more simulations starting from fully developed conditions at $Wi = 50$ and suddenly decreasing $Wi$ to $Wi = (5,6,7,8)$. The corresponding results (green markers in Fig.~\ref{fig:Secondary_flow_strength}b) almost perfectly overlap with the measures of the order parameter made by ramping up the Weissenberg number from rest (red markers). Note that, as the selected order parameter involves a temporal average, a perfect superposition cannot be expected. However, after cross-checking the time series, we remark that the resulting chaotic dynamics observed by ramping up (red markers) and ramping down (green markers) in $Wi$ lead to chaotic secondary flows with the same amplitude in $u_r$. We can therefore safely claim that our numerical simulations demonstrate, up to their computational precision, that the instability is compatible with a linear supercritical nature, as it does not admit any hysteresis loop. This is quantitatively illustrated in Table~\ref{tab:Instability}, where the values of the order-parameter versus $Wi$ are provided for both ramping-up (symbol $\uparrow$) and ramping-down (symbol $\downarrow$) protocols. An important remark is now in order. Since no time-marching simulation or experiment can exclude that an instability is subcritical, as they cannot approach the critical threshold due to the critical slowing down, a backward bifurcation with a small hysteresis loop could always be hiding in the (small but finite) interval $Wi_c \pm \delta Wi$ that experiments and time-marching simulations are limited to. We emphasize, however, that even if the bifurcation has a very narrow subcritical hysteresis cycle that we could not detect, such a local feature remains confined to $Wi \in (1\pm 0.1)\times Wi_c $ and does not limit the validity of the results presented in this study.

In summary, taking into account also the detailed linear stability analysis(LSA) reported in~\cite{Beneitez2025,Tej2025}, 
we conclude that the present instability 
could be due to: (i) a linear mechanism for a weakly two-dimensional Taylor-Couette flow, (ii) non-modal transient growth of small, yet not infinitesimal, perturbations, (iii) a nonlinear mechanism, or (iv) a linear mechanism induced by 
polymer diffusivity (i.e. polymer diffusive instability, or PDI)~\cite{Beneitez2025,Tej2025}. The first three hypotheses are tentative explanations for the instability we document, 
which is not supposed to occur in the one-dimensional inertialess, 
viscoelastic Taylor-Couette flow under infinitesimal perturbations. Assessing the validity of such hypotheses would require further detailed investigations that are out of the scope of 
the present work. 
As for 
hypothesis (iv), 
while this may be plausible, we note that quantitatively different predictions obtained in that framework can be found in the literature. 
The 
LSA by Beneitez et al. \cite{Beneitez2025} 
predicts a critical threshold 
(see light-blue shading in Fig.~\ref{fig:Secondary_flow_strength}b, $Wi^{Beneitez\ et\ al.}_c=0.58\pm 0.17$)
nearly one order of magnitude smaller than the 
one identified 
here, namely $Wi_c \approx 9.5\times Wi^{Beneitez\ et\ al.}_c$ (red dashed line in Fig.~\ref{fig:Secondary_flow_strength}b), still leaving room also for hypotheses (i-iii). 
On the other hand, the LSA by Tej et al. \cite{Tej2025} 
predicts a critical Weissenberg number (see orange shading in Fig.~\ref{fig:Secondary_flow_strength}b, $Wi^{Tej\ et\ al.}_c=3.6\pm 0.3$) compatible with ours, i.e. $Wi_c \approx 1.5\times Wi^{Tej\ et\ al.}_c$. Such a result then 
suggests that PDI may indeed be relevant for the present study. 
Figure 9 by Tej et al. \cite{Tej2025} 
further shows that the critical Weissenberg number does not seem to be significantly affected by the value of the diffusion coefficient,
which would explain why 
our results in terms of $Wi_c$ are grid-converged (see Fig.~\ref{fig:Phi_c}b). 
The same work~\cite{Tej2025} indicates that PDI occurs for nondimensional polymer diffusion coefficients 
$\mathcal{D}=D_p\lambda/(R_o-R_i)^2\in[10^{-5},\ 10^{-2}]$, where $D_p$ is the dimensional polymeric diffusivity. 
In a typical water solution, the latter is of the order of $D_p \approx 10^{-12}\,\text{m}^2/\text{s}$ (see~\cite{dubief2023elasto}). 
Given a characteristic relaxation time $\lambda \in [1, 10]\,\text{s}$ bracketing the critical onset threshold $\lambda_c = Wi_c/\Omega$ and assuming $\mathcal{D}=10^{-2}$ because it is compatible with the order of our estimated numerical diffusivity ($A_c|\nabla \bm{u}| \in [0.02, \ 1.64] \times 10^{-2}$ for the fine grid and $A_c|\nabla \bm{u}|\in [0.01, \ 1.23] \times 10^{-2}$ for the extra-fine grid at $Wi \in [5, \ 6]$, where $A_c$ denotes the finite volume area; see also~\cite{Beneitez2025} for such an estimate of the diffusivity) and critical wavenumber $m_c\in \lbrace 2,3\rbrace$
(see Fig.~\ref{fig:Phi_c}c), 
a \textit{physically-consistent} PDI  would then require a characteristic gap width of the order of tens to hundreds of microns. 
Interestingly, the upper-bound of 
this interval is not far from some of the gap-width values adopted in experiments~\cite{steinberg2021}.
Such an order-of-magnitude estimate then seems to suggest that the implicit diffusion of our numerical solver 
could be comparable to the \textit{physical} polymer diffusivity
(neglected in the Oldroyd-B model formulation). 

\begin{table}[!h]
    \centering
    \medskip
    \begin{ruledtabular}
    \begin{tabular}{lcc | lcc}
    $Wi$ &$\Phi_{u_r}\vert_{r=0.351}^{Wi\uparrow}$ & $\Phi_{u_r}\vert_{r=0.351}^{Wi\downarrow}$ & $Wi$& $\Phi_{u_r}\vert_{r=0.351}^{Wi\uparrow}$& $\Phi_{u_r}\vert_{r=0.351}^{Wi\downarrow}$\\
    \bottomrule
    1    & $<10^{-4}$ &           & 8    & $7.57\times 10^{-3}$& $8.33\times 10^{-3}$ \\
    4.5  & $<10^{-4}$ &                        & 9    & $8.39\times 10^{-3}$\\
    5    & $<10^{-4}$ & $<10^{-4}$         & 10   & $8.84\times 10^{-3}$&\\
    5.5  & $1.32\times 10^{-4}$& &12.5 & $9.62\times 10^{-3}$&\\
    6    & $4.25\times 10^{-3}$&{$4.78\times 10^{-3}$} & 25   & $1.11\times 10^{-2}$&\\
    6.5  & $5.99\times 10^{-3}$& &50   & $1.15\times 10^{-2}$&\\
    7    & $6.48\times 10^{-3}$&{$7.02\times 10^{-3}$}& 100  & $9.93\times 10^{-3}$& \\
    \end{tabular}
    \caption{Order parameter $\Phi_{u_r}$ evaluated at $r=0.351$ by ramping-up the control parameter $Wi$ from rest ($\Phi_{u_r}\vert_{r=0.351}^{Wi\uparrow}$) and ramping-down from fully developed conditions at $Wi = 50$ ($\Phi_{u_r}\vert_{r=0.351}^{Wi\downarrow}$).}\vspace{-0.5cm}
    \label{tab:Instability}
    \end{ruledtabular}
\end{table} 

We end this section by remarking that the elastic instability of this 2D flow was also discussed in two previous recent studies~\cite{buel_elastic_2018,vanBuel_2024} relying on time-marching simulations. In the former~\cite{buel_elastic_2018}, several simulations were performed starting from rest, for increasing Weissenberg numbers. The authors concluded that when $Re\ll 1$, $\eta = 1/4$, and $\beta = 0.6$ the flow becomes supercritically unstable at $Wi_c = 9.99 \pm 0.08$. This finding was then revised in the second study~\cite{vanBuel_2024}, showing some evidence of a hysteresis loop for $5.5 \leq Wi \leq 16$, when ramping up and down across the critical conditions. Hence, it was proposed that the instability has a subcritical character. Both these results are in disagreement with our findings, according to which the transition is plausibly supercritical and occurs at $Wi_c \simeq 5.5$. We therefore tried to identify the source of this inconsistency, which cannot be attributed to the implementation of the numerical solver, since all the concerned studies employ the same toolbox (RheoTool). The grid used in Refs.~\cite{buel_elastic_2018,vanBuel_2024} is our coarsest one, i.e. $N_r \times N_{\phi} = 100 \times 120$, for which we do find a transition at $Wi_c\approx 10$ for the simulations that ramp-up in $Wi$ from rest (see Appendix A). However, upon a refinement of the grid, our critical Weissenberg number decreases down to $Wi_c \simeq 5.5$, as assessed by our results for $N_r \times N_{\phi} = 200 \times 360$. Moreover, ramping-down from fully-developed supercritical conditions at $Wi = 50$, we do not observe any indication of a hysteresis loop when descending to $Wi=(5,6,7,8)$. We therefore conclude that the discrepancy with the results of Refs.~\cite{buel_elastic_2018,vanBuel_2024} should be due to the spatial resolution adopted in those studies which does not seem sufficient to fully resolve the dynamics. 
\subsection{Integral flow properties in a regime of fully developed elastic turbulence}\label{sec:global}

The four highest Weissenberg-number cases ($Wi = 12.5, 25, 50, 100$) strongly deviate from the theoretical bifurcation prediction resulting from a local linearization around $Wi_c$. This reflects the importance gained by the nonlinearities of the Oldroyd-B dynamics in highly supercritical conditions. As commonly done for inertial turbulence, we start the statistical characterization of the strongly supercritical ($>2\times Wi_c$) flow by considering global quantities, such as the kinetic ($E_k$) and elastic ($E_e$) energies. In nondimensional form, scaling them by $\mu \Omega$, these energies are respectively defined as
\begin{equation}
    E_k  = \frac{Re}{2S}\int_{S} \bm{u}^{2} dS, \quad E_e   = \frac{1-\beta}{Wi} \frac{1}{S} \int_{S} 
    tr(\bm{C}-\bm{I}) dS,
    \label{eqn:global_energy}
\end{equation}
where $S = \pi (1-\eta^2)$ denotes the two-dimensional domain area.

\begin{figure}[htbp]
\centering
    \includegraphics[width=0.675\linewidth]{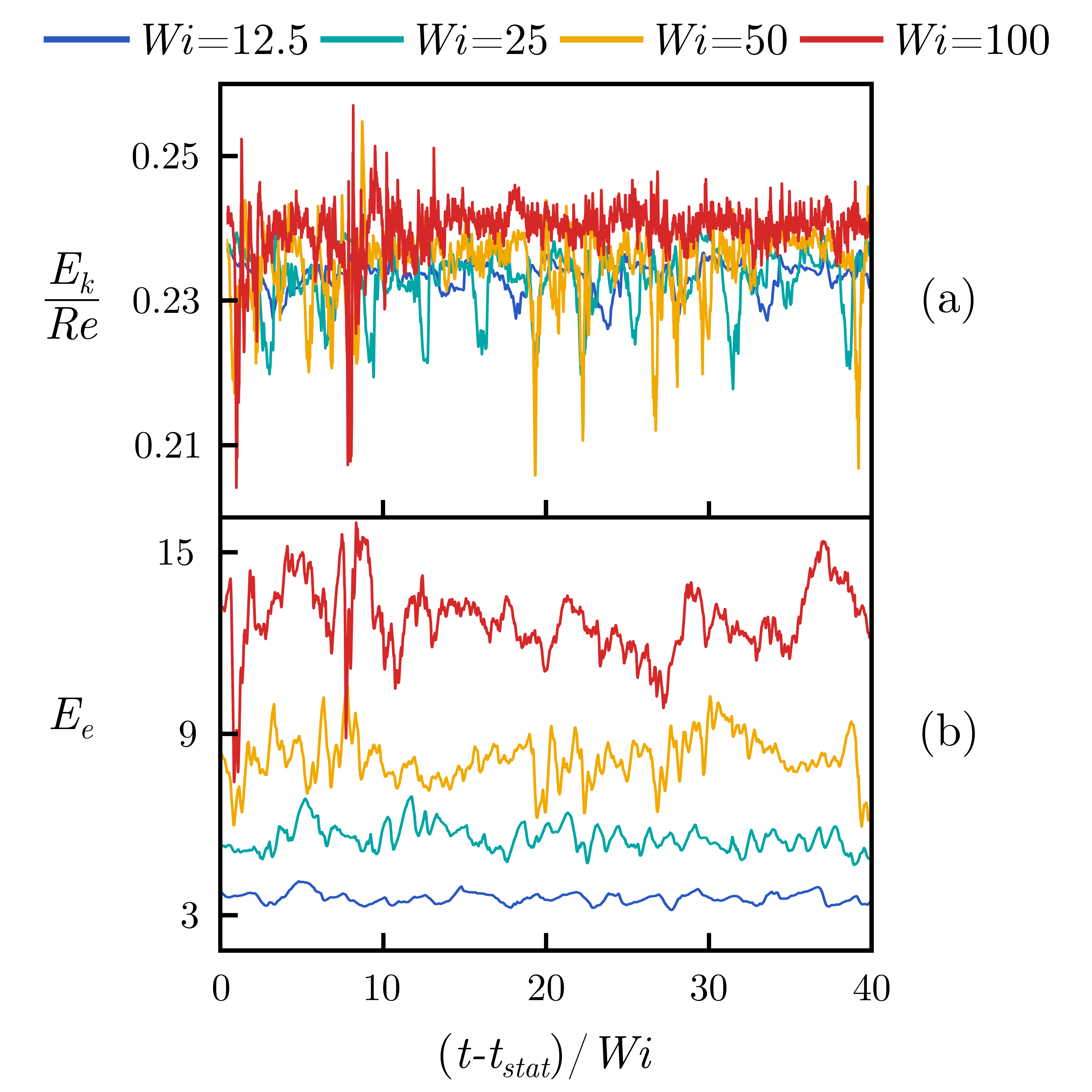}
\caption{
Kinetic energy $E_k$ divided by $Re$ (top) and elastic energy $E_e$ (bottom) versus time, for $Wi = 12.5, 25, 50, 100$. The transient to reach the statistically steady state ($t< t_{stat} = 600$) was removed, the durations of the shown time series correspond to $40 Wi$ in nondimensional time units.}
\label{fig:KE_EE}
\end{figure}

The magnitude of the kinetic energy rescaled with the Reynolds number \cite{steinberg2021} is found to be $E_k/Re = \mathcal{O}(0.1)$, which indicates that  inertial effects are negligibly small. 
Although the kinetic energy is independent of $Wi$, the elastic energy depends on it in a way that we clarify in the following. We also note that significantly higher frequencies are observed in the kinetic energy compared to the elastic energy for all four Weissenberg numbers depicted in Fig.~\ref{fig:KE_EE}. This is understood as a result of the two response time scales for the momentum (kinetic energy) and polymeric stress (elastic energy) equations. Interpreting the control parameters as timescale ratios, the Reynolds number represents the ratio between viscous ($t_\nu = R_o^2 \rho/\mu$) and convective ($t_c = 1/\Omega$) time scales, i.e. $Re= t_\nu/t_c$. The Weissenberg number (here equivalent to the Deborah number) rather denotes the ratio between the polymeric ($t_p = \lambda$) and the convective time scales, i.e. $Wi = t_p/t_c$. Therefore, considering that $Re = 2.5 \times 10^{-4}$ and $Wi \in [12.5,\ 100]$, their ratio $Wi/Re = t_p/t_\nu \in [5, 40] \times 10^4 \gg 1$ suggests that viscous diffusion responds much faster than polymeric stress. Moreover, the momentum equation is forced by the divergence of the polymeric stress $\nabla\cdot\bm{\tau}_p$, which implies smaller scales and higher frequencies than in $\bm{\tau}_p$. Hence, we interpret the higher frequencies observed in the kinetic energy as the result of a faster response ($Wi/Re \gg$ 1) to a faster forcing ($\nabla\cdot\bm{\tau}_p$). As a last note on Fig.\ \ref{fig:KE_EE}, we anticipate that although the simulations at $Wi = 100$ are overall sound, their highest frequencies may not be fully exempt from numerical artifacts.

\begin{figure}[htbp]
\vspace{-0.75cm}
\centering
\includegraphics[width=0.675\linewidth]{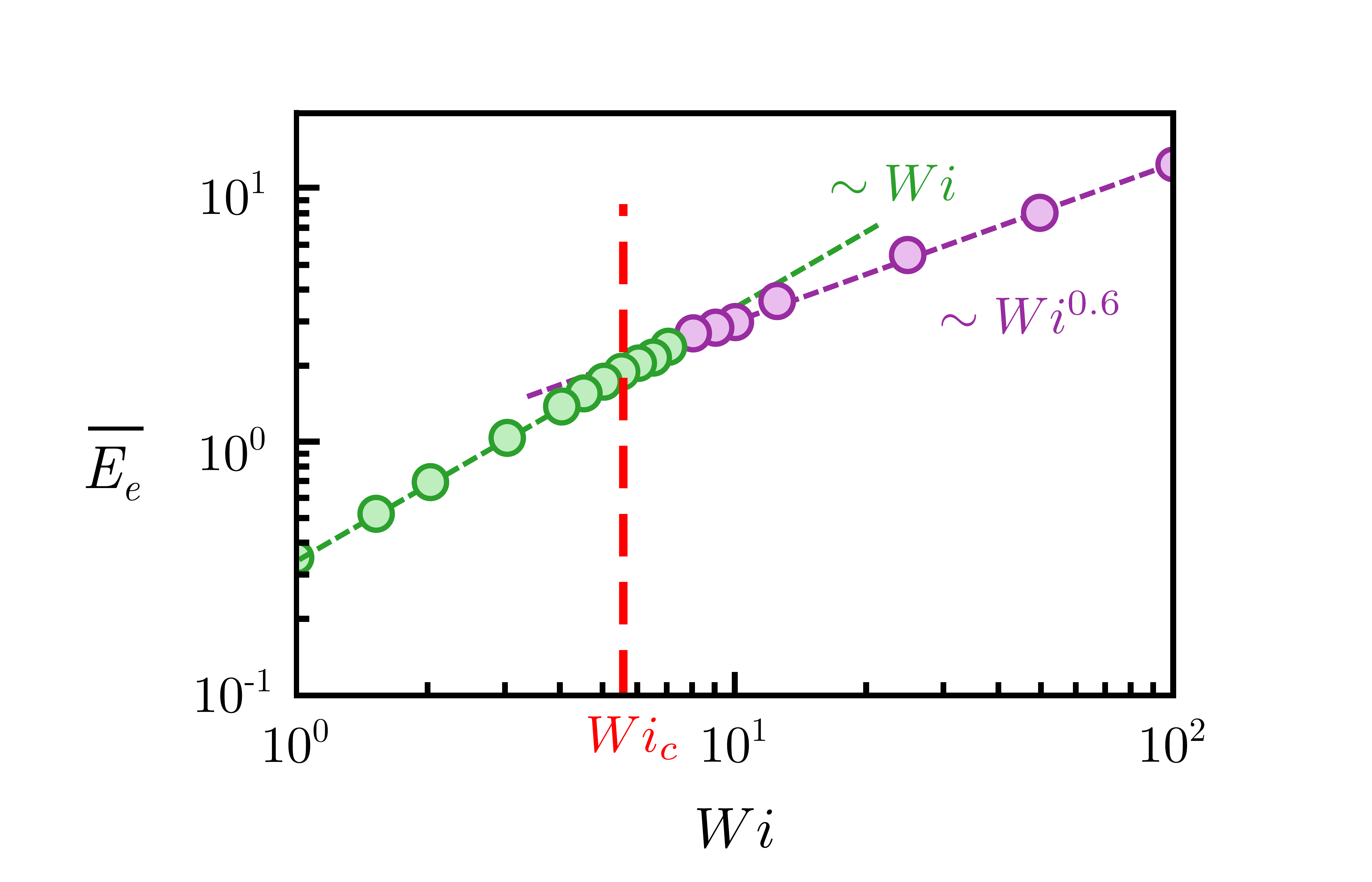}\vspace{-0.25cm}
\caption{Time-averaged elastic energy $\overline{E_e}$ as a function of $Wi$. The green markers for $Wi<1.3\times Wi_c$ match the subcritical steady linear scaling $\overline{E_e}\sim Wi$. The purple markers and the corresponding power-law fit, $\overline{E_{e}}\sim Wi^{0.6}$ for $Wi > 1.3\times Wi_c$, denote the supercritical scaling.}
\label{fig:EE_scaling}
\end{figure}

To investigate the dependence of $E_e$ on $Wi$, we consider the temporal averages $\overline{E_e}$ 
(over the duration 40$Wi$, equivalent to $40\lambda$) of the time series shown in Fig.~\ref{fig:KE_EE}. As shown in Fig.~\ref{fig:EE_scaling}, we find that $\overline{E_e}\sim Wi$ for $Wi < 1.3\times Wi_c$, which indicates that in the subcritical and weakly supercritical regime, the elastic-energy scaling is dominated by the laminar-regime behavior inherited from the basic state. Indeed, using the definition of $\bm{\tau}_p$ and the basic-state solution in Eq.~(\ref{eqn:basic_stress}), one has that $\overline{E_e} \sim Wi$. On the other hand, supercritical features determine the elastic-energy scaling for $Wi > 1.3\times Wi_c$, which is found to be $\overline{E_e}\sim{Wi}^{\gamma}$, with $\gamma=0.6$. 

As this second scaling applies starting from weakly supercritical conditions, we speculate that it is due to a redistribution of the mean extra stress caused by the nonstationarity of the flow. In other terms, part of the elastic energy is dissipated by the flow pulsations and does not contribute to the time-averaged elastic energy. As a result, the scaling found for $Wi > 1.3\times Wi_c$ has an exponent lower than the one for $Wi < 1.3\times Wi_c$. We conclude, therefore, that the supercritical scaling $\overline{E_e}\sim{Wi}^{0.6}$ is due to the flow instability rather than being a qualifying feature of the fully developed elastic-turbulence regime.

To start characterizing in more detail the structure of the flow, we now look at the fluctuating kinetic energy in the radial and azimuthal directions. Specifically, we examine the root-mean-square (rms) velocity components $u^{rms}_{r}$ and $u^{rms}_{\phi}$, which provide information on the degree of isotropy of the velocity field. These are computed as $u^{rms}_* = \langle \left(u_* - {\langle u_* \rangle_\phi} \right)^2\rangle_S^{1/2}$, where $\langle \cdot \rangle_S$ denotes a surface average and $*$ means either $r$ or $\phi$. Note that the definition of the latter quantity is motivated by the nonhomogeneity of the flow along $r$. The rms velocity should account for fluctuations with respect to the mean flow. Averaging over $\phi$ and keeping the dependence on $r$ when computing the quadratic deviation then allows us to get rid of the mean flow profile to focus on the flow fluctuations.

\begin{figure}[htbp]
    \includegraphics[width=0.625\linewidth]{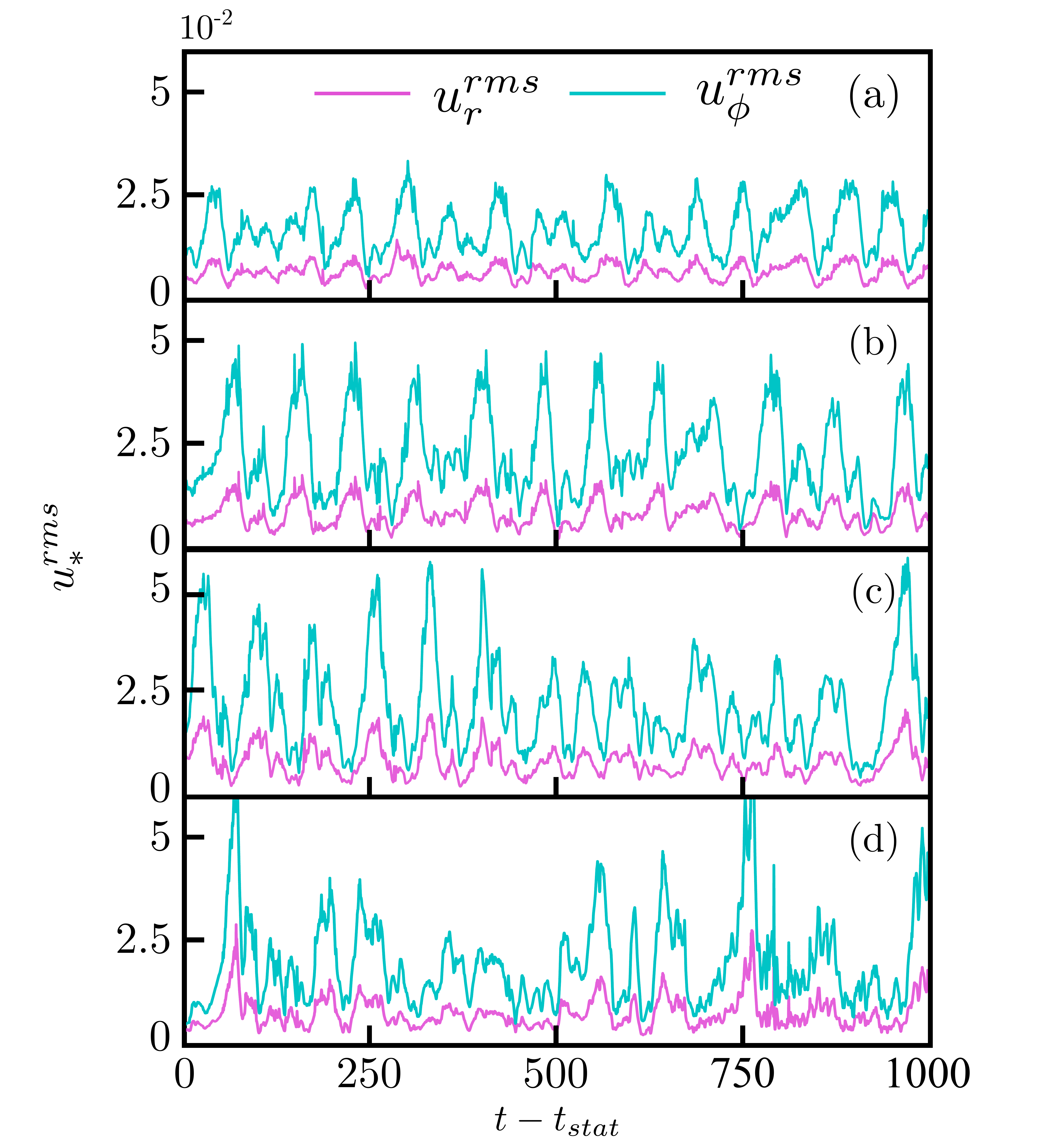}
    \caption{Time behavior of the rms velocity components $u^{rms}_{r}$ and $u^{rms}_{\phi}$ for $Wi = 12.5$ (a), 25 (b), 50 (c), 100 (d) increasing from top to bottom. The transient to reach the statistically steady state ($t < t_{stat} = 600$) was removed.}
    \label{fig:RMS}
\end{figure}

Figure~\ref{fig:RMS} reports the time series of $u^{rms}_{r}$ and $u^{rms}_{\phi}$ for a duration of $t - t_{stat} \in [0,\ 1000]$ in the statistically steady state. It is here evident that as $Wi$ grows, both fluctuations have a more irregular behavior. Independently of the Weissenberg number, however, we can also note the presence, in both $r$ and $\phi$ directions, of slow oscillations at a frequency $f_\text{rms} \approx 0.1$ of that corresponding to the rotation rate of the outer cylinder, $\Omega/(2\pi)$. We anticipate that this will have implications for the interpretation of energy spectra. After averaging over time, we find that the ratio $\overline{u_\phi^{rms}} / \overline{u_r^{rms}}$ is $2.7 \pm 0.1$ (for all $Wi$ values), which shows that fluctuations are larger in the azimuthal direction than in the radial one. As a function of $Wi$, such a ratio only shows a slight tendency to increase.
\subsection{Boundary layer for strongly supercritical conditions}\label{sec:boundary_layer}
From the visualizations of the secondary flow in Fig.~\ref{fig:Secondary_flow_strength}, it is apparent that the instability develops near the inner cylinder. This is expected, since, as in other curvilinear shear flows, the mechanism of the linear instability identified in Sec.~\ref{sec:instability} is driven by the hoop stress associated with the curvature of streamlines~\cite{shaqfeh1996purely,Pakdel_1996,larson_1990,Morozov_2005,steinberg2021}, which increases when approaching the inner wall. The decay of perturbations towards the outer cylinder should then imply the formation of a boundary layer, inside which elastic turbulence develops~\cite{groisman_2000,Burghelea_2006,Burghelea_2007,steinberg2021}. Sufficiently far from this region, a state close to the laminar one should instead be recovered, similarly to what happens in the presence of an inertial-turbulence boundary layer fed by linear instabilities~\cite{BLT_Schlichting}. The presence of such an elastic boundary layer would in turn imply that the energy injected in the system must redistribute across scales differently inside and outside this dynamically more active region. Experimental support for this observation seems to be provided by the two-sloped power spectra (in the 3D Taylor-Couette setup), arising from the spatial variations of the secondary-flow structure, reported in Ref.~\cite{groisman_2004}. Therefore, for an in-depth understanding of our viscoelastic 2D Taylor-Couette system, it is necessary to identify and characterize the boundary-layer region. 

Let us preliminarily remark that, more precisely, in the present case two boundary-layer widths can be determined, one based on the intensity of the elastic stresses at the origin of the instability, and one based on the secondary-flow strength. To estimate them, in both cases we proceed as follows. We first take the azimuthal average of the considered observable in the supercritical state. Thereafter we take the absolute value and average in time. Finally, we subtract the basic-state expression of the corresponding observable. The boundary-layer width is then defined as the distance such that this difference is $5\%$ of the maximum value of the temporal average of the given observable, once averaged over $\phi$ and in absolute value:
\begin{equation}
    r_{BL,\dagger}:\ \frac{|\overline{\ |\langle\star(r,\phi,t)\rangle_\phi |\ }- \star^0(r)|}{ \max(\overline{\ |\langle\star(r,\phi,t)\rangle_\phi| \ })}  = 0.05,
    \label{eqn:rBL}
\end{equation}
where $\star$ stands for $tr(\bm{\tau}_{p})$ or $u_r$ in the case of the elastic ($\dagger = e$) or the kinetic ($\dagger = k$) boundary layer, respectively. The regions $\eta\leq r\leq r_{BL,\dagger}$ then correspond to the two boundary layers. Their thicknesses $r_{BL,\dagger}$ essentially represent the radial distance from the inner wall beyond which the given observable relaxes to its basic state solution within a relative tolerance of 5\%. Note that taking the absolute value before the time average at the numerator of \eqref{eqn:rBL} is necessary for our estimate of the kinetic boundary layer. In fact, as we employ the radial component of velocity (which can be either positive or negative) to quantify the secondary-flow intensity, if we carry out the time-average omitting the absolute value, contributions of different sign would cancel out (to numerical accuracy), forbidding the estimation of the magnitude of $u_r$. 
\begin{figure}[p]
    \centering \vspace{-1.0cm}
       \includegraphics[width=0.65\linewidth]{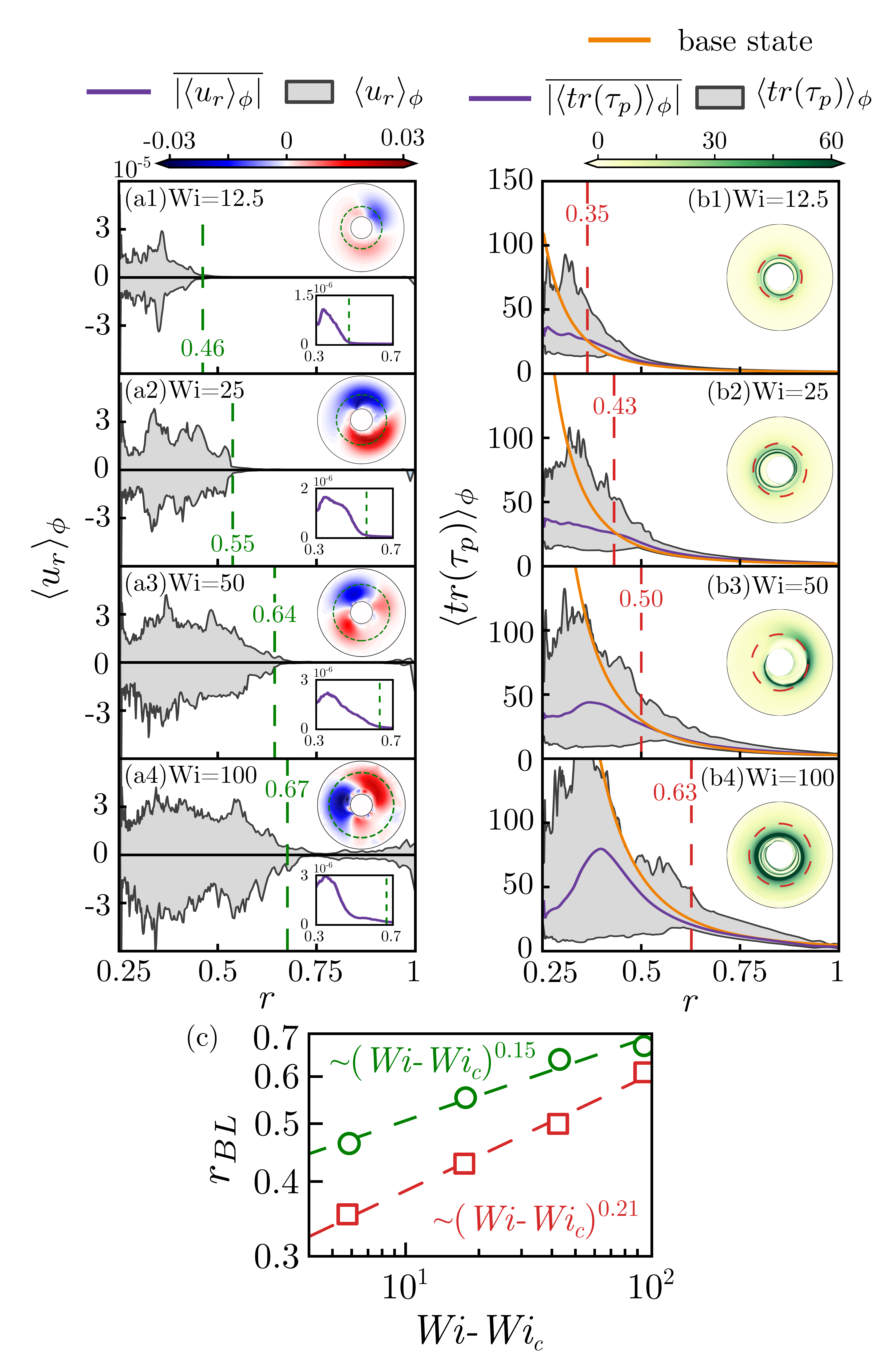}\vspace{-0.5cm}
    \caption{
    Azimuthally-averaged radial velocity $\langle u_r \rangle_\phi$ (a1--a4) and trace of the elastic stress $\langle tr(\bm{\tau}_p)\rangle_\phi$ (b1--b4) versus the distance from the inner wall, for $Wi = 12.5, 25, 50, 100$ from top to bottom. For both fields, the temporal averages of the absolute value of $\langle tr(\bm{\tau}_p)\rangle_\phi$ and $\langle u_r \rangle_\phi$, over a duration of $40 \, Wi$ in the statistically steady state, are shown by 
    purple solid lines, with $\overline{ \ \vert\langle u_r \rangle_\phi\vert \ }$ shown separately in the bottom right insets of panels (a1--a4); the variability over time is represented by the shaded grey regions; the basic-state solution for $tr(\bm{\tau}_p^0)$ is shown by the orange solid lines, while it is null for $u_r^0$. The elastic ($r_{BL,e}$) and kinetic ($r_{BL,k}$) boundary-layer widths are shown by vertical dashed red and green lines, respectively. In both columns, instantaneous snapshots of the considered fields are also shown, together with the corresponding boundary layer thicknesses (dashed circular lines). The bottom panel (c) shows the $Wi$ dependence of $r_{BL,e}$ (red line, rectangles) and $r_{BL,k}$ (green line, circles).} 
    \label{fig:bl}
\end{figure}

The results for the four strongly supercritical Weissenberg numbers, i.e. $Wi = 12.5$, 25, 50, 100, are reported in Fig.~\ref{fig:bl}. In this figure, $Wi$ grows from top to bottom, the measurements of $u_r$ and $tr(\bm{\tau}_p)$ are presented in the left and right columns, respectively. The purple lines denote the temporal average of the absolute-valued azimuthal averages of the signals (i.e. $\overline{\ |\langle\star(r,\phi,t)\rangle_\phi |\ }$) computed in the statistically steady state (for a duration of $40 \, Wi$). In both columns, the light-grey shading indicates the regions delimited by the envelopes of the azimuthal averages for each $r$, i.e. $\max_t (\langle\star\rangle_\phi)$ and $\min_t (\langle\star\rangle_\phi)$. The basic state is denoted by the orange line, whose proximity to the purple line defines the radial extent of each boundary layer according to Eq.~(\ref{eqn:rBL}) (also indicated by red-dashed lines for $r_{BL,e}$ and green-dashed lines for $r_{BL,k}$). The appropriateness of the boundary-layer thicknesses' estimates can be further visually appreciated by inspecting the (instantaneous) snapshots of the $u_r$ (left) and $tr(\bm{\tau}_p)$ (right) fields shown as insets in Fig.~\ref{fig:bl}, where the dashed lines indicate $r_{BL,k}$ and $r_{BL,e}$, respectively. Here, it is evident that the production of elastic energy, and hence of the secondary flow, is almost entirely confined to the regions $\eta\leq r\leq r_{BL,\dagger}$. The same information, extended over $40$ polymer relaxation times, is conveyed by the envelopes of $tr(\bm{\tau}_p)$ and $u_r$, which tend to shrink around the laminar-state solution for $r> r_{\text{BL},\dagger}$. Interestingly, this further highlights the co-existence of the two flow regimes (turbulent-like for $\eta\leq r\leq r_{BL,\dagger}$ and laminar for $r> r_{BL,\dagger}$) within our viscoelastic system for strongly supercritical conditions, i.e. $Wi > 2\times Wi_c$.

When the Weissenberg number is increased, both boundary layers become thicker (Fig.~\ref{fig:bl}c), in agreement with the intuition of dynamical nonlinearities becoming more important. Moreover, we find that $r_{BL,k}>r_{BL,e}$, independently of $Wi$, and that both display power-law growth with $Wi$. Indeed, the data quite closely follow the behaviors $r_{BL,\dagger} \sim (Wi - Wi_c)^{\xi_\dagger}$, with $Wi_c = 5.525$, and $\xi_e = 0.21$ and $\xi_k = 0.15$ from best fits (see Fig.~\ref{fig:bl}c). Based on these results, it appears reasonable to expect that, as we shall show, the statistical properties of elastic turbulence (as quantified, e.g., by power spectra) change in correspondence with different scales, depending on the (kinetic or elastic) nature of the observable considered.
\subsection{Energy spectra}\label{sec:spectra}
As documented in the previous sections, for large enough $Wi$ the flow attains an irregular, turbulent-like behavior. The statistical properties of such flow states can be quantified by energy spectra. Most experimental studies of elastic turbulence focus on kinetic energy spectra in the frequency ($f$) domain. This is because point measurements of time series are easier to perform than measurements of long time series of full flow fields with high spatial and temporal resolutions. For the physical interpretation of such frequency spectra, Taylor's frozen-field hypothesis is routinely used. 
Typically, these studies report power-law scaling $\mathcal{E}_k(f) \sim f^{-\zeta_k}$ with $\zeta_k >3$ (see~\cite{steinberg2021} for a review). 
Note, however, that in the 3D Taylor-Couette flow experimental spectra show two regions of power-law decay with different exponents, $\zeta_k \simeq 1.1$ at low frequencies and $\zeta_k \simeq 2.2$ at large ones~\cite{groisman_2004,steinberg2021}. Similar spectra are found at mid gap in numerical work~\cite{Liu_Khomami_2013}, though at $Re \approx 10$, which does not exclude possible inertial effects, and with velocity probability distributions differing from the experimental ones. Using Taylor's hypothesis, which in spite of some limitations was found to hold for second-order statistics (as, e.g., spectra) in shear-flow experiments~\cite{Burghelea_2005,Burghelea_2007} and simulations~\cite{garg2021statistical}, one then has that spectra in the wavenumber domain behave as $\mathcal{E}_k(m) \sim m^{-\sigma_k}$ with $\sigma_k=\zeta_k$ (with $m=|\bm{m}|$ the wavenumber modulus). Such steep spectra ($\sigma_k>3$), pointing to a smooth flow, mainly dominated by large-scale modes, find some justification in the theory of elastic turbulence developed in homogeneous isotropic conditions~\cite{fouxon_spectra_2003}. 

To the best of our knowledge, there exists no consensus on the universality of the scaling exponents $\sigma_k$, $\sigma_e$. While experimental studies seem to converge on $\sigma_k \approx 3.5$, numerical studies seem to suggest a larger value, $\sigma_k \approx 4$~\cite{nguyen2016small,lellep2024purely,singh2024intermittency}. A further prediction of the theory proposed in \cite{fouxon_spectra_2003}, which has been much less investigated, concerns the elastic energy spectra. These are also expected to display a power-law decay, $\mathcal{E}_e(m) \sim m^{-\sigma_e}$, with $\sigma_e=\sigma_k-2$~\cite{fouxon_spectra_2003,steinberg2021}. We stress, however, that the lack of the statistical symmetries (homogeneity and isotropy) assumed by the theory may lead to deviations from these behaviors, as also reported in previous numerical studies in $2D$ setups~\cite{gupta2019effect,canossi2020elastic}. The numerical studies \cite{nguyen2016small,lellep2024purely,singh2024intermittency} indicate values of $\sigma_e$ in the range $\left[1.5, 2.3\right]$. In Ref. \cite{singh2024intermittency}, Singh and coworkers have derived a different relationship between the scaling exponents of the kinetic and elastic spectra, $\sigma_k=2\sigma_e+1$, which they validate via direct numerical simulations (DNS) performed in a homogeneous Arnold-Beltrami-Childress 
flow. While discussing the spectra, it is also important to note that, unlike kinetic energy spectra, elastic energy spectra are not directly accessible in experiments, because the available methods to measure elastic stresses are very limited. Elastic spectra can be inferred somewhat indirectly by measurements of pressure fluctuations or torques (e.g. in rheometric flows). This then makes DNS a unique tool to test the related theoretical predictions.

In the present system, the intensity of turbulent-like dynamics clearly varies with the radial distance $r$, being higher close to the inner cylinder and decaying further away from it (see Sec.~\ref{sec:boundary_layer}). Therefore, we measure spectra, of both kinetic and elastic energy, at different positions $r$ in the gap. The frequency (or temporal) and wavenumber (or spatial) spectra are then computed as follows:
\begin{align}
\label{eq:spectra_def}
    &\mathcal{E}_k(f,r) = \langle|\hat{\bm{u}}'_f|^2\rangle_\phi, \ & &\mathcal{E}_e(f,r) = 
    \langle |\hat{\ell}'_f|^2
    \rangle_\phi,\\
     &\mathcal{E}_k(m,r) = \overline{|\hat{\bm{u}}'_m|^2},& &\mathcal{E}_e(m,r) = \overline{|\hat{\ell}'_m|^2}, \nonumber
\end{align}
where $\ell \equiv \sqrt{tr(\bm{\tau}_p)}$, the hat denotes a Fourier transform, the subscript $f$ indicates that the latter is performed on a time series at a given $(r,\phi)$ location, the subscript $m$ indicates a transform over the azimuthal variable $\phi$ at given $r$ and $t$, and prime indicates a fluctuation, computed  with respect to the spatial average for $\hat{\bm{u}}'_f$ and $\hat{\ell}'_f$, and with respect to the azimuthal average for $\hat{\bm{u}}'_m$ and $\hat{\ell}'_m$.

\begin{figure}[t]
    \centering
       \includegraphics[width=0.675\linewidth]{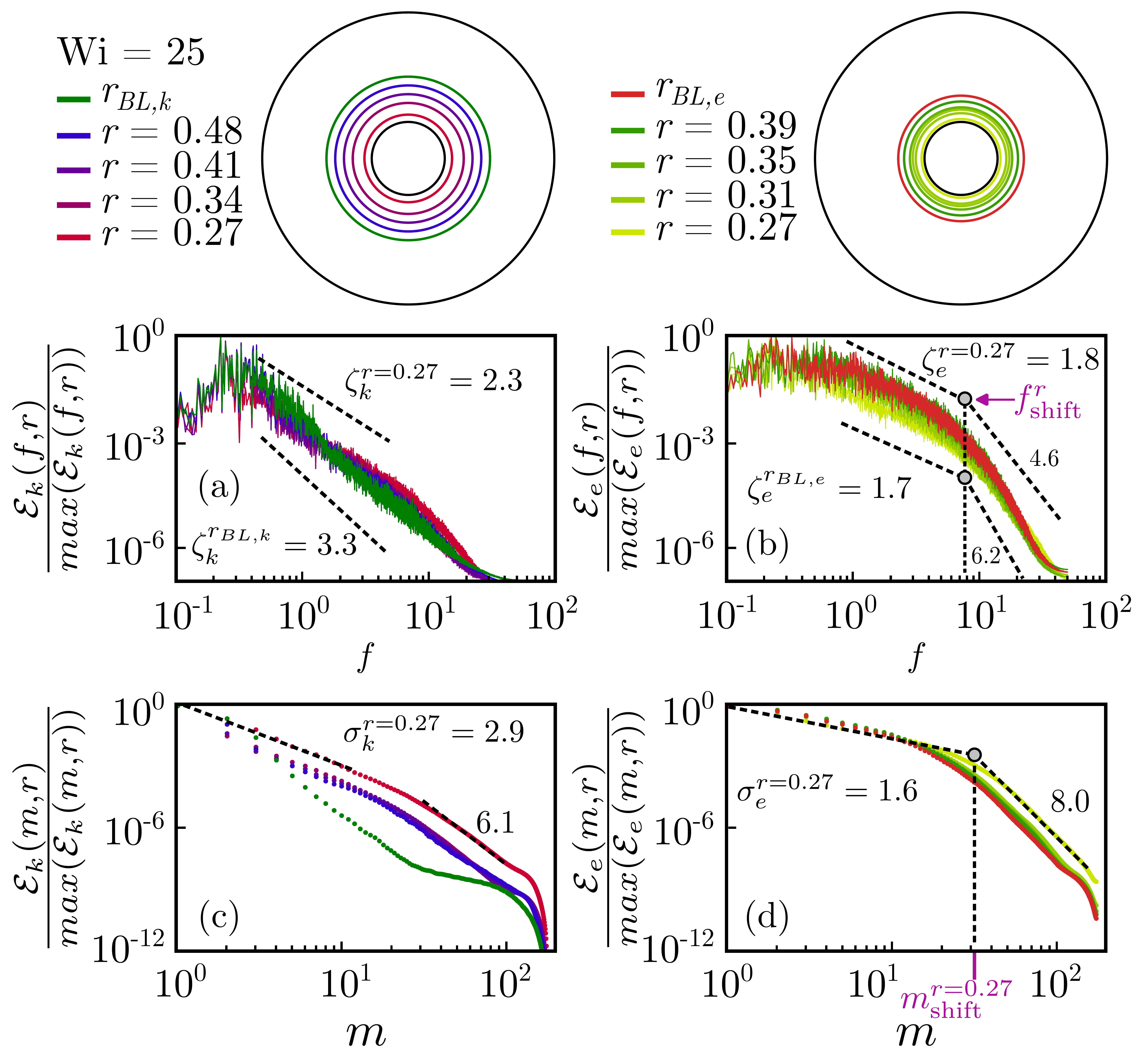}
    \caption{Temporal (a,b) and spatial (c,d) spectra of kinetic (a,c) and elastic (b,d) energy at different radial positions within the respective boundary layers $\eta \leq r \leq r_{BL,e}$ and $\eta \leq r \leq r_{BL,k}$, for $Wi=25$. The specific values of $r$ chosen are shown, for both cases, by the contours in the topmost panels. All spectra are normalized by their maximum value to ease comparison. The dashed lines indicate power-law behaviors (of exponents $\zeta_{e,k}$ and $\sigma_{e,k}$). The exponents are obtained from a best fit in the low-frequency/wavenumber subranges. At higher frequencies/wavenumbers, due to the limited scaling range, these lines are intended as guides for the eye.
    In elastic spectra, clearer changes of slope can be identified at the frequency $f_\text{shift}$ or the wavenumber $m_\text{shift}$. Note that the frequency $f$ in (a) and (b) is the nondimensional one (i.e. the dimensional one scaled by $\Omega/2\pi$).}
    \label{fig:spectra_Wi25}
\end{figure}

We start the discussion on spectra by considering those of elastic energy, which are directly related to the stress generating elastic turbulence. These are shown in Figs.~\ref{fig:spectra_Wi25}b and \ref{fig:spectra_Wi25}d at different radial positions in the corresponding boundary layer, $r \leq r_{BL,e}$, for $Wi=25$. The latter is here taken as a reference Weissenberg number for turbulent-like states. A discussion on the $Wi$ dependence of these results is provided at the end of this section and in Appendix B. The corresponding kinetic energy spectra are also shown, for  $r \leq r_{BL,k}$ and $Wi=25$, in Figs.~\ref{fig:spectra_Wi25}a and \ref{fig:spectra_Wi25}c. 

Irrespective of the value of $r$, the temporal spectra $\mathcal{E}_e(f,r)$ quite clearly display two distinct decay subranges. At low frequencies ($0.5 \lesssim f \lesssim 6$) characteristic of slower and larger flow structures, we find $\mathcal{E}_e(f,r) \sim f^{-\zeta_e}$ with $1.7 \leq \zeta_e \leq 1.8$. At larger frequencies ($10 \lesssim f \lesssim 30$), the scaling range is shorter than a decade, therefore the value of the exponent $\zeta_e$ (possibly $4.6 \leq \zeta_e \leq 6.2$) is more uncertain and should be taken with some caution. Note that similar considerations apply to the high-wavenumber subrange of spatial spectra. 
However, it is clear that such temporal spectrum gets considerably steeper at the largest frequencies. The low frequencies are related to elastic turbulence, while the high frequencies are produced in the very proximity of the inner cylinder and they get readily dissipated by friction at the wall. The switch between these two behaviors is observed to occur at frequencies slightly smaller than $10$, namely $f_\text{shift}\in [6,\ 8]$, almost independent of the Weissenberg number, as confirmed by the results for the other strongly supercritical cases ($Wi=12.5,50,100$), reported in Appendix B. The most energetic flow features are thus associated with frequencies from $\approx 0.1$ to $\approx 10$ of the driving frequency $\Omega/(2\pi)$. Based on the results shown in Fig.~\ref{fig:RMS}, the typical slow frequency of the velocity fluctuations ($f_\text{rms} \approx 0.1$) seems to be responsible for it, in combination with a similar slope-switching occurring in the spatial spectrum at a wavenumber $m_\text{shift}$, giving the periodicity in $\phi$ of the smallest sustainable flow structures. We speculate that the frequency $f_\text{shift}$ is associated with the transport, over a duration $1/f_\text{rms}$ of the smallest energetic structures (with $m \approx m_\text{shift}$), which would give a temporal periodicity of such flow patterns with frequency $f_\text{rms} \, m_\text{shift}$. Using the measured values $m_\text{shift} \approx 30$ and $f_\text{rms} \approx 0.1$, we find reasonable agreement with this expectation, as we get $f_\text{rms} m_\text{shift} \approx 3$, which is of the same order as $f_\text{shift} \approx 6$.

As anticipated above, spatial spectra also display a change of 
behavior, in correspondence with a wavenumber $m_\text{shift}=O(10)$, only slightly varying with $Wi$ between $20$ and $40$. Specifically, at small wavenumbers we find that $\mathcal{E}_e(m,r) \sim m^{-\sigma_e}$ with $\sigma_e \approx 1.6$, while at larger ones $\sigma_e \approx 8.0$. Once again, we associate the first, larger-scale subrange with elastic turbulence. The much steeper decay of the spectrum in the second subrange instead suggests efficient viscous energy dissipation at small scales.
The same picture is returned by the spectra of the other supercritical cases (see Appendix B). To rationalize the crossover between these two behaviors of $\mathcal{E}_e(m,r)$, we reason geometrically. At a distance $r$ from the inner wall, the maximum size of eddies should be of order $r$. At the same time, and independently of $r$, in the azimuthal direction the angular periodicity of the flow patterns selected by the instability imposes a lower bound on the typical eddy size. This periodicity is expressed by an azimuthal wavenumber, which we argue to be $m_\text{shift}$. At even smaller scales, instead, we expect dissipative effects to dominate. In particular, owing to the azimuthal periodicity of the Taylor-Couette flow, the secondary structures produced are cellular flows whose aspect ratio is prone to be of $\mathcal{O}(1)$. This implies that the leading wavenumbers in radial and azimuthal directions are comparable. Near the inner cylinder, this also implies that we can allocate several small, roughly square cells. Hence, the corresponding wavenumbers $m\gg 1$ are high near the inner cylinder. In turn, this implies that they will undergo a strong dissipation due to their leading Fourier mode: if $u_r \sim \sin(m\phi)$ and $m\gg 1$, the azimuthal dissipation $\mathcal{D_\phi}\simeq\partial_{\phi\phi}u_r \sim -m^2\sin(m\phi)$ is significantly more important near the inner cylinder than in the bulk. Moreover, as the secondary structures appear in the form of roughly square cells, $\mathcal{D_\phi}\approx\mathcal{D}_{r}$, the dissipation in both coordinate directions is promoted near the inner cylinder. Such interpretation is visually confirmed in our simulations, where the highest wavenumbers are dominant near the inner wall and their lifetime is significantly shorter than the propagation time for the main secondary flow circulation. This should then imply that the dynamically active eddies have sizes in between the lengthscale $\Lambda_\text{shift}=2\pi \, r/m_\text{shift}$ and $r$. Moving away from the wall, the driving mechanism of the instability weakens, due to the reduced streamline curvature, so that it appears reasonable that the full dynamics can only sustain larger, more energetic scales.  

\begin{figure}[!h]
\vspace{-0.3cm}
    \centering
    \includegraphics[width=0.675\linewidth]{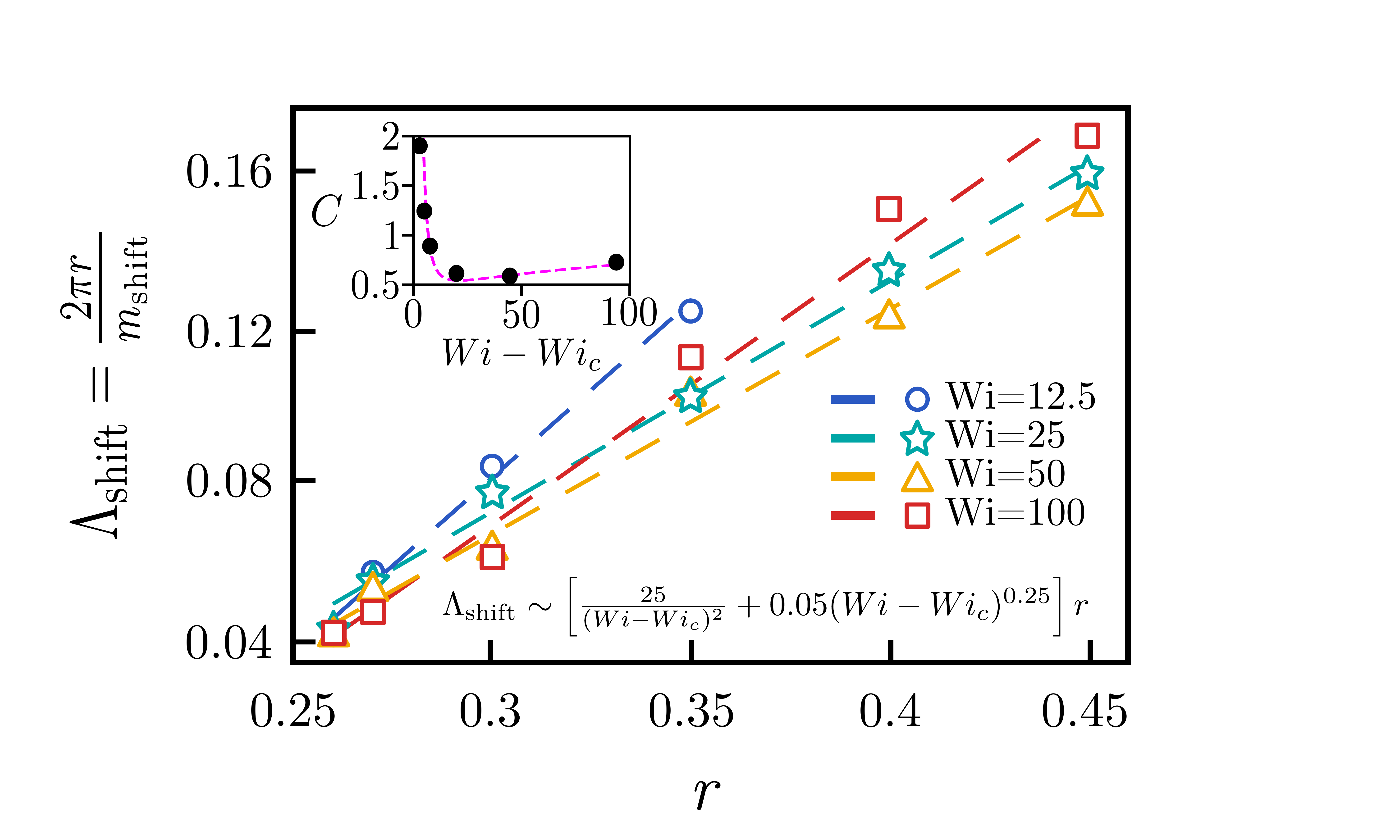}
    \caption{Crossover length scale $\Lambda_\text{shift}$ fitted by a linear function in $r$, $\Lambda_\text{shift} \sim C(Wi,\ Wi_c) \times r$. The inset shows that the prefactor of this linear scaling is well described by $C(Wi,Wi_c) = \left[ \frac{25}{(Wi-Wi_c)^2}+0.05(Wi-Wi_c)^{1/4} \right]$, for $2.5 \leq Wi-Wi_c \leq 94.5$.}
    \label{fig:shifting_wavenumber}
\end{figure}

As a first approximation, we further expect $\Lambda_\text{shift}$ to weakly depend on the  Weissenberg number if $Wi \gg Wi_c$, as our geometric argument does not rely on any scaling with $Wi$, but solely on the co-existence of elastic turbulence and of viscosity-dominated scales. Our data support such a picture, as shown in Fig.~\ref{fig:shifting_wavenumber}, where a linear scaling with $r$ is found, $\Lambda_\text{shift} \sim C(Wi,\ Wi_c) \times r$, with a prefactor depending on the actual and critical Weissenberg numbers as elucidated below. Our interpretation of $\Lambda_\text{shift}$ implies that for wavelengths larger than $\Lambda_\text{shift}$, elastic turbulence can be sustained. Hence, $C(Wi,\ Wi_c)$ should become singular at $Wi=Wi_c$, leading to $\Lambda_\text{shift}\rightarrow \infty$ for $Wi \rightarrow Wi_c$. In other words, this means that no shift to an elastic-turbulence scaling should occur if the elastic instability does not take place. On the other hand, for large Weissenberg numbers, we expect that an asymptotic scaling exists with $(Wi - Wi_c)^{\gamma_\text{shift}}$. Once again, these expectations are in good agreement with our data (see inset of Fig.~\ref{fig:shifting_wavenumber}), which show that $C$ diverges as $(Wi-Wi_c)^{-2}$ for $Wi \simeq Wi_c$, while it has a weak sublinear dependence (with $\gamma_\text{shift}=1/4$) on $Wi$ for $Wi\gg Wi_c$. 

\begin{figure}[t]
\centering
\includegraphics[width=0.675\linewidth]{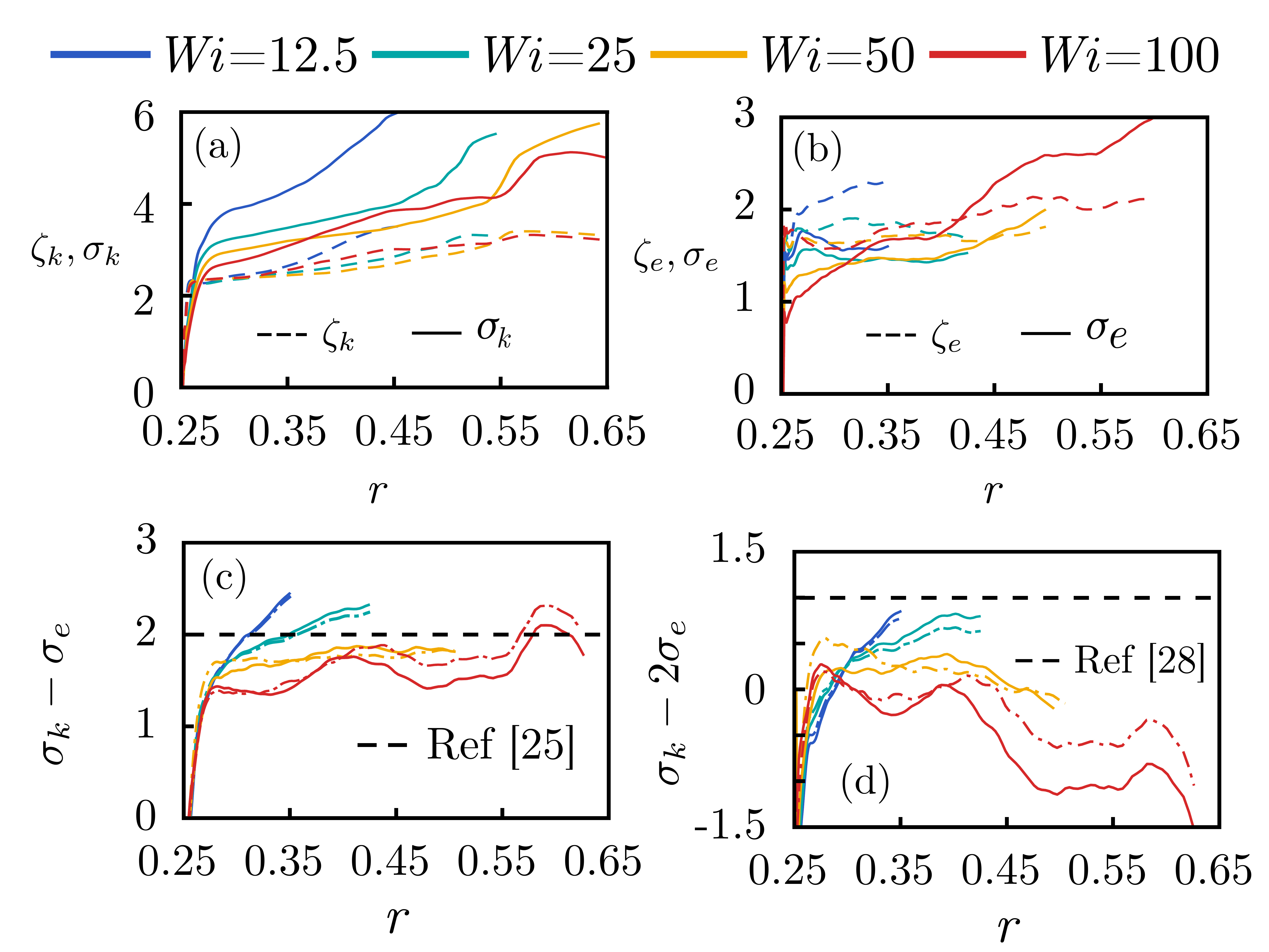}
\caption{
Radial dependency of the fitted slopes of temporal and spatial spectra for $Wi=12.5,\ 25,\ 50,\ 100$ of kinetic energy (a) and elastic energy (b). For spatial spectra, the difference of the kinetic and elastic energy spectrum exponents of the respective power-law fits, $\sigma_k-\sigma_e$ is also shown in (c) as a function of $r$. In this plot, the black dashed line represents the theoretical prediction $\sigma_k-\sigma_e=2$~\cite{fouxon_spectra_2003}. 
The comparison with the prediction $\sigma_k-2\sigma_e=1$~\cite{singh2024intermittency} is shown in (d). 
Note that in (c) and (d) the exponents $\sigma_e$ are extracted from spectra computed both as in Eq.~(\ref{eq:spectra_def}) (solid lines) and as in~\cite{singh2024intermittency} (dashed lines); more details can be found in the text.}

\label{fig:coefficient_all}
\end{figure}

As for kinetic energy spectra (Fig.~\ref{fig:spectra_Wi25}a and Fig.~\ref{fig:spectra_Wi25}c for $Wi=25$, and Appendix B for other values of $Wi$), we find that the temporal ones, $\mathcal{E}_k(f,r)$, are characterized by a power-law behavior, of exponent $2.3<\zeta_k<3.3$, at relatively low frequencies. 
Nevertheless, also here a tendency of the spectrum to become steeper for $f \gtrsim 10$ is observed. The extensions of these subranges are comparable with those found from elastic energy spectra. The values of $\zeta_k$ mentioned above appear compatible with those in the majority of elastic turbulence experiments, but somewhat larger than those found in some previous numerical and experimental studies in the 3D Taylor-Couette flow~\cite{groisman_2004,Liu_Khomami_2013}. Sufficiently far from the inner boundary, and independently of $Wi$ (see Fig.~\ref{fig:coefficient_all}a), however, they are not far from $\zeta_k \approx 3$, as also recently reported from experimental spectra of azimuthal velocity fluctuations at mid gap~\cite{zhang2025experimental}. Spatial spectra, $\mathcal{E}_k(m,r)$, display a clearer transition, which becomes more evident closer to the inner cylinder, between a low-wavenumber range, where $\sigma_k \simeq 2.9$, and a high-wavenumber one, where possibly $\sigma_k \approx 6.1$. The wavenumber at which such a shift occurs is not far from the value of $m_\text{shift}$ obtained for the elastic energy spectra. The robustness of this observation when changing the Weissenberg number (see Appendix B) suggests that the shift of decay behavior detected in $\mathcal{E}_k(m,r)$ is a byproduct of the one manifesting in $\mathcal{E}_e(m,r)$. We remark that the values of $\sigma_k$ for $m<m_\text{shift}$ are larger than those (at mid gap) from 3D DNS~\cite{song2023self}. Moreover, we stress that well outside the boundary layer ($r>r_{BL,k}$) kinetic energy spectra are much steeper (not shown), consistently with the decay of elastic turbulence in this outer region (see Fig.~\ref{fig:bl}).
Finally, the larger values of the spectral slopes (in both the frequency and wavenumber domains) with respect to those found in experiments~\cite{groisman_2004} and 3D DNS~\cite{song2023self} suggest that three-dimensionality promotes energetic small scales, while in 2D the flow is smoother and more dominated by larger structures.

A more global view of the radial dependency of the exponents of the temporal and spatial spectra (for both elastic and kinetic energy) is provided in Fig.~\ref{fig:coefficient_all} for all the supercritical cases explored. Note that here we refer to the scaling subranges $f<5$ and $m<m_\text{shift}$ only. Moreover, the different curves are plotted only for distances $\eta \leq r \leq r_{BL,e}$ and $\eta \leq r \leq r_{BL,k}$, meaning inside the boundary layers, where spectra can be reasonably described by power-law functions. Clearly, in the region immediately adjacent to the inner wall, viscosity dominates and one cannot expect power-law scaling. As it can be seen, all the exponents associated with the elastic energy spectra (Fig.~\ref{fig:coefficient_all}b) are almost constant (roughly between $1$ and $2$), except for $Wi=100$, with $\sigma_e$ further increasing (outwards in $r$) to $\approx 3$. As for kinetic energy spectra, the exponents $\zeta_k$ weakly vary with $r$ in the range $2 \lesssim \zeta_k \lesssim 4$. The exponents $\sigma_k$ behave quite similarly, but also exhibit fast growth close to the edge of the boundary layer. Such an increase appears dynamically consistent, as the boundary layer delimits the energetically active turbulent region, beyond which velocity fluctuations decay.

Finally, in homogeneous isotropic conditions, it has been theoretically predicted that the exponents of the power-law decays (in the wavenumber domain) of kinetic and elastic energy are related by $\sigma_k-\sigma_e=2$~\cite{fouxon_spectra_2003,steinberg2021}. Our determination of $\sigma_e$ and $\sigma_k$ allows testing this prediction as a function of the distance $r$. For the different Weissenberg numbers considered, our data display non-negligible deviations from this prediction (Fig.~\ref{fig:coefficient_all}c).
The comparison with the more recent theoretical prediction $\sigma_k - 2\sigma_e = 1$~\cite{singh2024intermittency} is also presented in Fig.~\ref{fig:coefficient_all}d; 
 this relation was also considered by Ref.~\cite{serafini2026} using a Lagrangian polymer model in an idealized, triperiodic homogeneous flow.
Note, however, that  in~\cite{singh2024intermittency}, the elastic energy spectrum is not defined as in Eq.~(\ref{eq:spectra_def}) but rather as $\mathcal{E}_e(k) \sim \langle \hat{B}_{\gamma\beta}(\mathbf{m})\hat{B}_{\beta\gamma}(-\mathbf{m}) \rangle$ with $k = |\mathbf{m}|$, 
and where $\bm{B}$ is the unique square root of the conformation tensor 
($\bm{C}=\bm{B}\bm{B}^T$). 
To assess the robustness of our conclusions against these 
slightly different definitions of 
the elastic energy spectrum, Figs.~\ref{fig:coefficient_all}c and d employ the exponents $\sigma_e$ obtained via 
the above definition~\cite{singh2024intermittency} (dashed-dotted lines) alongside our original 
one (solid lines). 
For $Wi\leq 50$, the variations in $\sigma_k - \sigma_e$ and $\sigma_k - 2\sigma_e$ remain relatively insensitive to the 
definition of $\mathcal{E}_e(m)$, whereas a noticeable (yet minor) deviation emerges at $Wi=100$. Overall, the $\sigma_k - 2\sigma_e = 1$ 
relation appears less successful in capturing our numerical data (Fig.~\ref{fig:coefficient_all}d). The numerical results systematically deviate from this prediction across nearly the entire radial span. Summarizing, the discrepancy between numerical spectra and theoretical predictions obtained in homogeneous isotropic conditions underscores a breakdown of both the predicted scalings when applied to the present Taylor–Couette configuration, characterized by pronounced large-scale spatial nonhomogeneity.


We conclude this section by briefly commenting on the validity of Taylor's hypothesis in our system. Its use is typically justified when velocity fluctuations are small compared to the mean flow. This seems to be the case in the present simulations, as it can be understood from Fig.~\ref{fig:RMS}, showing that velocity fluctuations (as quantified by the rms components $u_r^{rms}$, $u_\phi^{rms}$) do not exceed $\approx 6\%$ of the driving velocity of the external cylinder (unity in nondimensional units), giving the typical intensity of the mean flow. Despite such a consideration, the values of the exponents of power-law decay of spatial spectra tend to be systematically larger than those of temporal ones, hinting at a breakdown of the hypothesis. Note that relevant differences between the exponents of spatial and temporal spectra have also been recently reported from simulations in different setups~\cite{TH_Rosti}. Here, a possible reason could be that, near the inner cylinder, we find that velocity fluctuations can be significant with respect to the local mean flow. Similar conclusions have been obtained in experiments in von Karman swirling flow~\cite{Burghelea_2005}. In that system, Taylor's hypothesis is found to break down in the inner flow region, due to large velocity fluctuations and small mean flow, while it seems to hold in the outer region. Further, more focused studies would be needed to fully assess the validity of this hypothesis in our system.

\section{Conclusions}\label{sec:concl}
We numerically investigated the 2D, inertialess, viscoelastic Taylor-Couette flow, focusing on its purely elastic instability and, subsequently, its elastic-turbulence regime. Previous work already reported on the interest of this system for elastic turbulence and mixing at low Reynolds number~\cite{buel_elastic_2018,vanBuel_2024}. Here we revisit the characterization of the elastic instability, addressing previously contrasting evidence, and highlight the nonhomogeneous character of the turbulent-like dynamics.

Using extensive DNS we find that, within numerical accuracy, the instability 
cannot be safely associated with the PDI reported by the LSA of \cite{Beneitez2025}, due to a one-order-of-magnitude discrepancy in the value of $Wi_c$, but it may be compatible with the PDI predicted by the LSA of \cite{Tej2025}, which gives closer quantitative agreement with the present results. The value of the numerical diffusivity that likely triggers it is found to be reasonably comparable to the physical diffusion coefficient of polymers in water solutions in microdevices. The critical Weissenberg number at which it occurs, measured from the behavior of the secondary flow intensity, is found to be somewhat smaller than previously anticipated~\cite{buel_elastic_2018}, namely $Wi_{c}\approx 5.5$. Moreover, in correspondence with $Wi \approx 1.3 \, Wi_c$, a transition can be detected from the behavior of the global elastic energy. Below $Wi \approx 1.3 \, Wi_c$, this scales linearly with $Wi$, as predicted by the laminar basic-state solution, while above $Wi \approx 1.3 \, Wi_c$ the scaling is sublinear, likely due to an increase of dissipation, associated with the onset of time dependency in the flow.

It is interesting to comment on the nature of the transition to elastic turbulence in the context of the coil-stretch transition undergone by polymers upon a gradual increase of $Wi$ in the flow. According to de Gennes~\cite{cstransition}, the coil-stretch transition is either subcritical if driven by a strain-dominated flow or supercritical when the driving flow is vorticity-dominated. Within this framework, our findings do not seem to suggest a subcritical transition as we find no significant trace of a hysteresis cycle. Therefore, our simulations align well with de Gennes' prediction ~\cite{cstransition} and the structure of the base flow, which is vorticity dominated. This indicates a direct relationship between the primary elastic instability and the coil-stretch transition responsible for the generation of elastic stresses in the flow.

For $Wi>Wi_c$, the main outcome of this study concerns the identification of a boundary-layer region, close to the inner cylinder wall, where nonlinear dynamics are mostly active. The kinetic boundary layer is found to be thicker than the elastic one, and the extents of both increase as power laws of $Wi$. This represents a novel feature of this system, which, to our knowledge, is still poorly documented in numerical work on elastic turbulence also in other setups. 
Although the presence of a boundary layer plays a crucial role in the development of elastic turbulence~\cite{steinberg2021,Belan_2018}, the lack of a direct method to measure the elastic stresses makes its experimental assessment rather difficult. Thus, the extent of the boundary layer is assessed experimentally via measurements of profiles of velocity fluctuations~\cite{Burghelea_2007,Soulies_etal_2017}.  However, as discussed throughout the text and illustrated in Fig.~\ref{fig:bl}c, the width of the boundary layer obtained from velocity distributions is systematically larger than the width determined from the stress distributions. This indicates that the experimental approach of inferring the elastic boundary layer via velocity measurements should be undertaken with care and perhaps revisited.

Further away towards the outer wall, elastic turbulence rapidly decays and the laminar, base flow is recovered. Turbulent-like states are then found to be characterized by strong nonhomogeneity, and moderate anisotropy. At leading order, the azimuthal velocity fluctuation $u'_\phi$ dominates over the radial fluctuation $u'_r$, with a root-mean-square ratio $u_\phi^{rms}/u_r^{rms}\sim2.7$. This ratio is neither large enough to justify a higher-order scale balance, nor sufficiently close to unity to account for $u_r'$ in the total kinetic energy scaling. This implies that the observed scalings are primarily enforced by $u'_\phi$. 

When the Weissenberg number is large enough, inside the active region, the dynamics are controlled by a range of (spatial and temporal) scales, as highlighted by power-law spectra of both kinetic and elastic energy. In the frequency domain, spectra quite gently steepen with the distance $r$ from the inner wall. The slope of the kinetic-energy spectrum is typically larger than the one found at mid gap in experiments and 3D DNS~\cite{groisman_1998,Liu_Khomami_2013}, varying between $\zeta_k \gtrsim 2$ (close to the inner cylinder) and $\zeta_k \lesssim 4$ (at larger distances). Independently of $r$, at large frequencies the spectrum tends to decay faster. This is better observed in the elastic-energy spectrum, which allows identifying a critical frequency for this transition that does not seem to depend on the Weissenberg number. In the wavenumber domain, spectral slopes increase with the distance $r$, first quite slowly, then more rapidly when $r$ approaches the boundary-layer thickness. Well inside the active region, those of kinetic-energy spectra attain values $2<\sigma_k<4$, to some extent compatible with the theoretical prediction in homogeneous isotropic conditions~\cite{fouxon_spectra_2003},  below a crossover wavenumber ($m_\text{shift}$). Beyond the latter, the spectrum falls off more rapidly. Interestingly, such a change of slope is even clearer in the spectrum of elastic energy, and a geometric argument suggests that, at leading order, the crossover lengthscale does not depend on $Wi$, as the physics involved at low and high wavenumbers are controlled by different energetic exchanges directly associated to the size of the eddies. 
At small scales ($m>m_\text{shift}$) the kinetic energy is readily dissipated by viscosity, while larger scales ($m<m_\text{shift}$) correspond to elastic-turbulence flow patterns. Relying on the database of our simulations for the four largest Weissenberg numbers, we proposed a quantitative fit for the shift wavelength, hoping to prompt further investigation in terms of scaling arguments. Concerning the relation between the exponents of the kinetic ($\sigma_k$) and elastic ($\sigma_e$) energy spectra, we find that this is not well captured by either of the two available predictions (obtained in homogeneous, isotropic conditions), $\sigma_k-\sigma_e=2$~\cite{fouxon_spectra_2003} and $\sigma_k-2\sigma_e=1$~\cite{singh2024intermittency}, which calls for further theoretical developments in nonhomogeneous flows. Finally, in the outer region outside the boundary layer, spatial spectra are much steeper, due to the rapid decay of fluctuations (with $r$). We also note that the data on spectra from our simulations  do not seem to be fully compatible with the validity of Taylor's hypothesis, which allows one to bridge results from the frequency to the wavenumber domain. This observation is in line with the relatively large intensity of velocity fluctuations, with respect to that of the mean flow near the inner cylinder, despite the fact that rms velocities, when averaged over the whole domain, are only about $0.5$ to $6\%$ of the driving velocity in our measurements. 

In future studies, it would be interesting to explore the dependence of the turbulent-like dynamics on the vertical-to-horizontal aspect ratio in quasi 2D to 3D setups. Moreover, addressing a detailed characterization of the mixing properties of this system, both from an Eulerian and a Lagrangian point of view, may reveal useful to assess the potential of nonhomogeneous elastic turbulence to achieve mixing at low Reynolds number in wall-bounded configurations.

\section*{Acknowledgments}
We wish to acknowledge the financial support of the China Scholarship Council (CSC) for Z. Hou (Student Number 202308070010) during his doctoral study in France.

\appendix
\subsection*{Appendix A: Grid and time step convergence tests}\label{app:GSI}
Four meshes have been tested for carrying out the grid convergence study: coarse mesh ($N_r \times N_{\phi} = 100 \times 120$), medium mesh ($N_r \times N_{\phi} = 150 \times 240$), fine mesh ($N_r \times N_{\phi} = 200 \times 360$) and extra fine mesh ($N_r \times N_{\phi} = 250 \times 480$). 

More details on the mesh size in the radial direction 
can be found in Table~\ref{tab:Mesh_size}. The simulations previously performed in Ref.~\cite{buel_elastic_2018} employ a mesh that corresponds to our coarse mesh, for which we reproduce qualitatively and quantitatively their results.

\begin{table}[!h]
    \centering
    \begin{tabular}{c|cccc}
       Mesh type       & $N_r$ & $N_\phi$ & $\min(\Delta r)$ & $\max(\Delta r)$ \\  \hline
       Coarse mesh     &   100    &    120      &  $1.979\times10^{-3}$                &           $1.880\times10^{-2}$       \\
       Medium mesh     &  150     &   240       &       $1.321\times10^{-3}$           &                 $1.255\times10^{-2}$ \\
       Fine mesh       &  200     &   360       &       $9.913\times10^{-4}$           &                 $9.417\times10^{-3}$ \\
       Extra-fine mesh &  250     &   480       &       $7.933\times10^{-4}$          &                  $7.537\times10^{-3}$\\ 
    \end{tabular}
    \caption{Characteristics of the finite-volume meshes tested in this study.}
    \label{tab:Mesh_size}
\end{table}

\begin{figure}[t]
    \centering
       \includegraphics[width=0.8\linewidth]{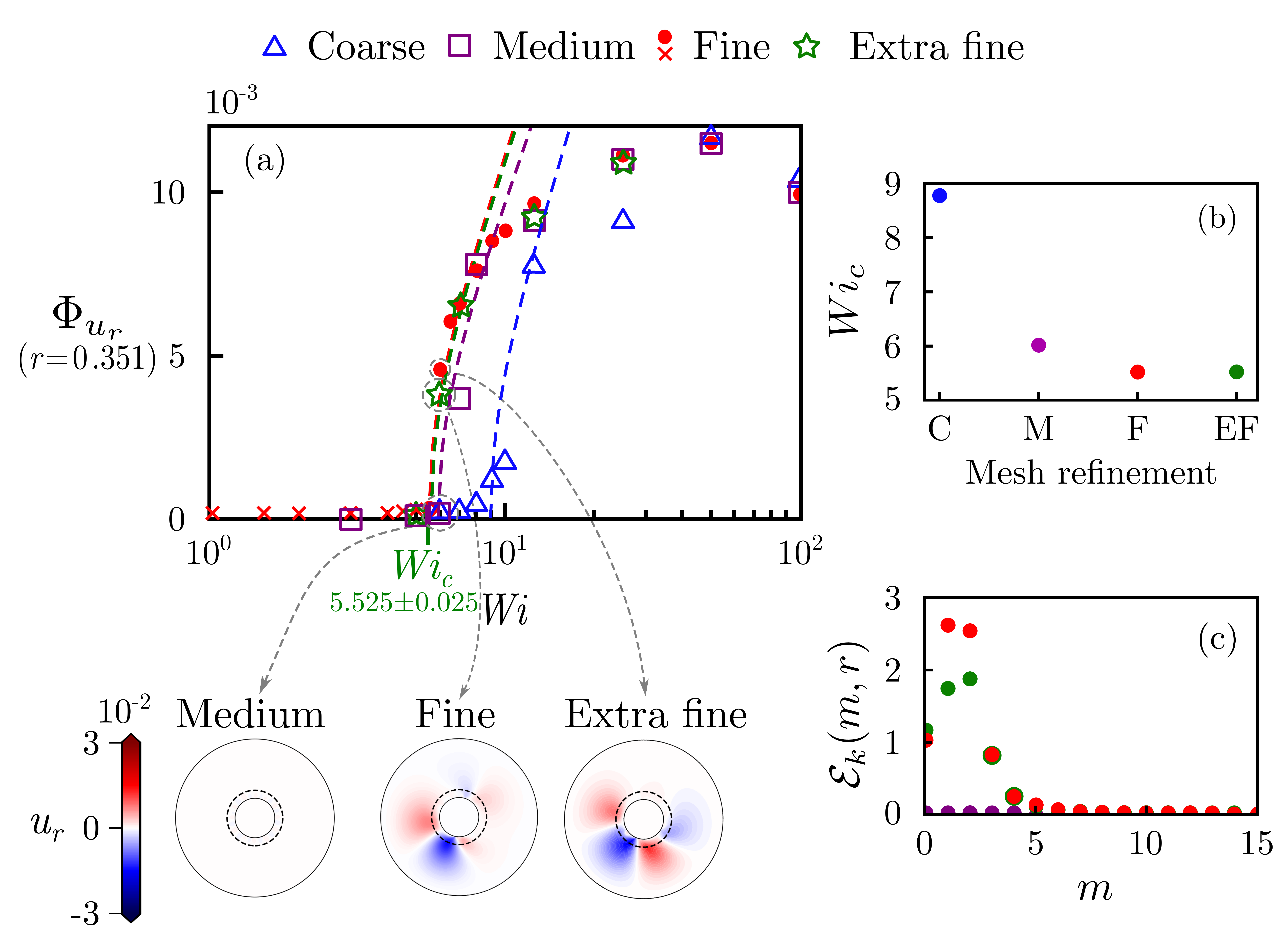}
    \caption{(a) Secondary flow strength, measured by the order parameter $\Phi_{u_r}$ evaluated at $r=0.351$, 
    versus $Wi$.
    The dashed lines denote fits with $(Wi - Wi_c)^{1/2}$. The red dots correspond to the results in the main text using fine mesh, which gives the critical Weissenberg number $Wi_c = 5.525\pm 0.025$. The blue triangles correspond to the results using coarse mesh. Two extra datasets, using medium mesh (purple squares) and extra fine mesh (green stars) are shown, which almost overlap with the results using fine mesh. 
    In the bottom left panels, the $u_r$ snapshots corresponding to different meshes at $Wi=6$ are shown.
    (b) Critical Weissenberg number $Wi_c$ for increasing mesh refinement (with C, M, F, EF standing for coarse, medium, fine, extra fine mesh, respectively).
    (c) Fourier spectra of the $u_r$ snapshots at $Wi=6$ for medium, fine and extra fine meshes at the given value of $r$} [with color coding as in panels (a) and (b)].
    
    \label{fig:Phi_c}
\end{figure}

Fig.~\ref{fig:Phi_c} presents a mesh convergence study on the secondary flow strength $\Phi_{u_r}$ at $r=0.351$ across the four considered meshes. The numerical results obtained using the coarse mesh (blue triangles) agree with those from Ref.~\cite{buel_elastic_2018}, as in this case we find $Wi_c \approx 10$, as reported by Ref.~\cite{buel_elastic_2018}. 
However, we obtain different results using the fine mesh adopted in the present work (Sec.~\ref{sec:results}). To assess the influence of the spatial grid resolution, we also computed $\Phi_{u_r}$ using data from the medium and extra fine meshes, obtaining results that closely overlap with those based on the fine mesh. 
This, together with the
convergence of $Wi_c$ 
with increasing mesh refinement, 
shown in Fig.~\ref{fig:Phi_c}b, 
provides support for the statement that our fine mesh gives converged results for determining
the bifurcation diagram of Sec.~\ref{sec:instability}. 
The visualizations of the radial velocity field at the bottom-left of Fig.~\ref{fig:Phi_c} further demonstrate the grid convergence of the fine grid. In addition, the Fourier spectrum $\mathcal{E}_k(m,r)$ in Fig.~\ref{fig:Phi_c}c shows that the secondary flow is dominated by the primary $m=1, 2, 3$ modes for the fine and extra-fine meshes. 

\begin{figure}[t]
    \centering
       \includegraphics[width=0.675\linewidth]{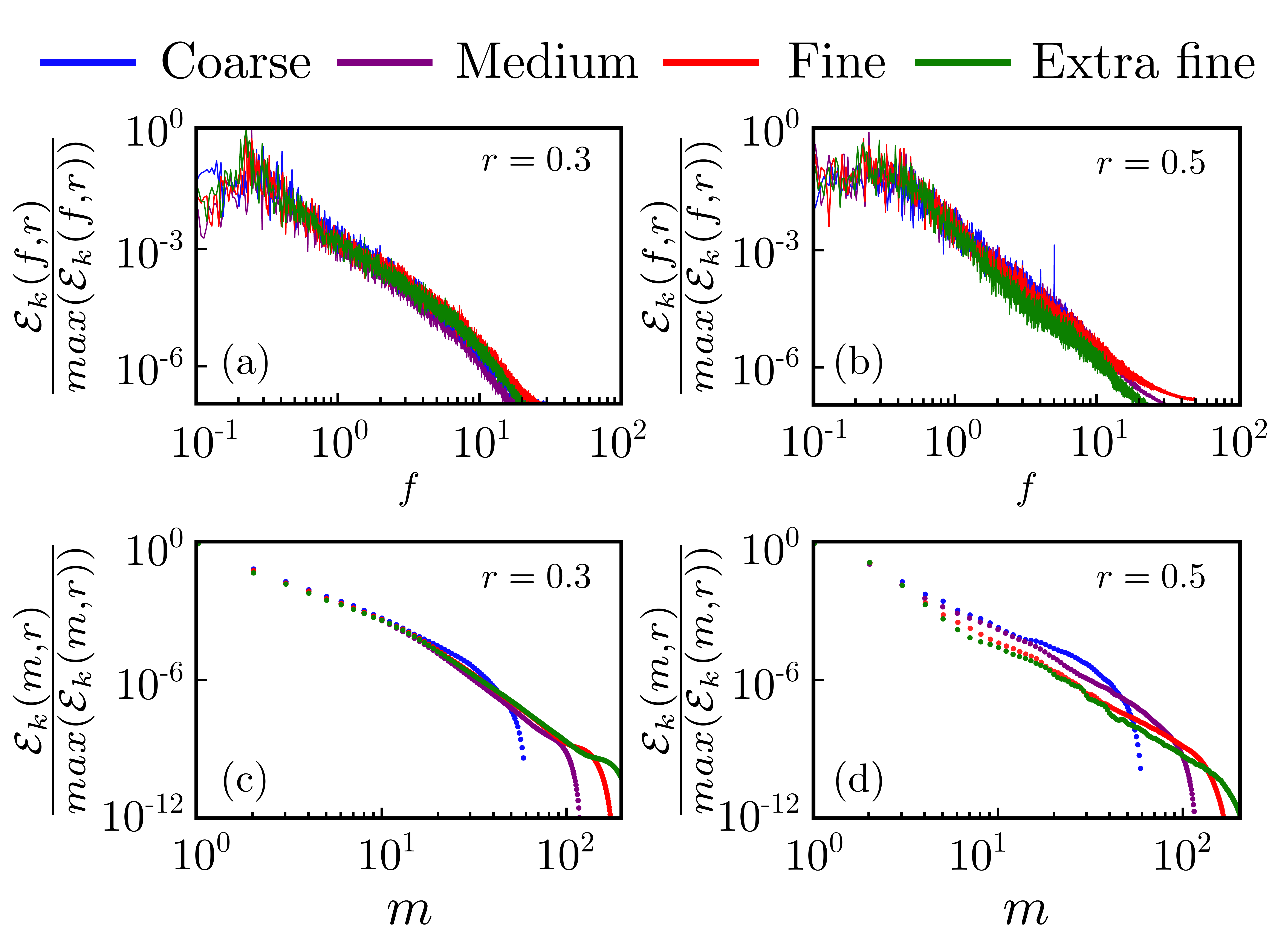}
    \caption{Grid-independence study through four levels of mesh refinement. The top line presents temporal kinetic energy spectra at two different radial positions ($r = 0.3$ and $0.5$), while the bottom one presents the corresponding spatial spectra, at $Wi=25$.}
    \label{fig:GCS} 
\end{figure}    
\begin{figure}[t]
    \centering
       \includegraphics[width=0.675\linewidth]{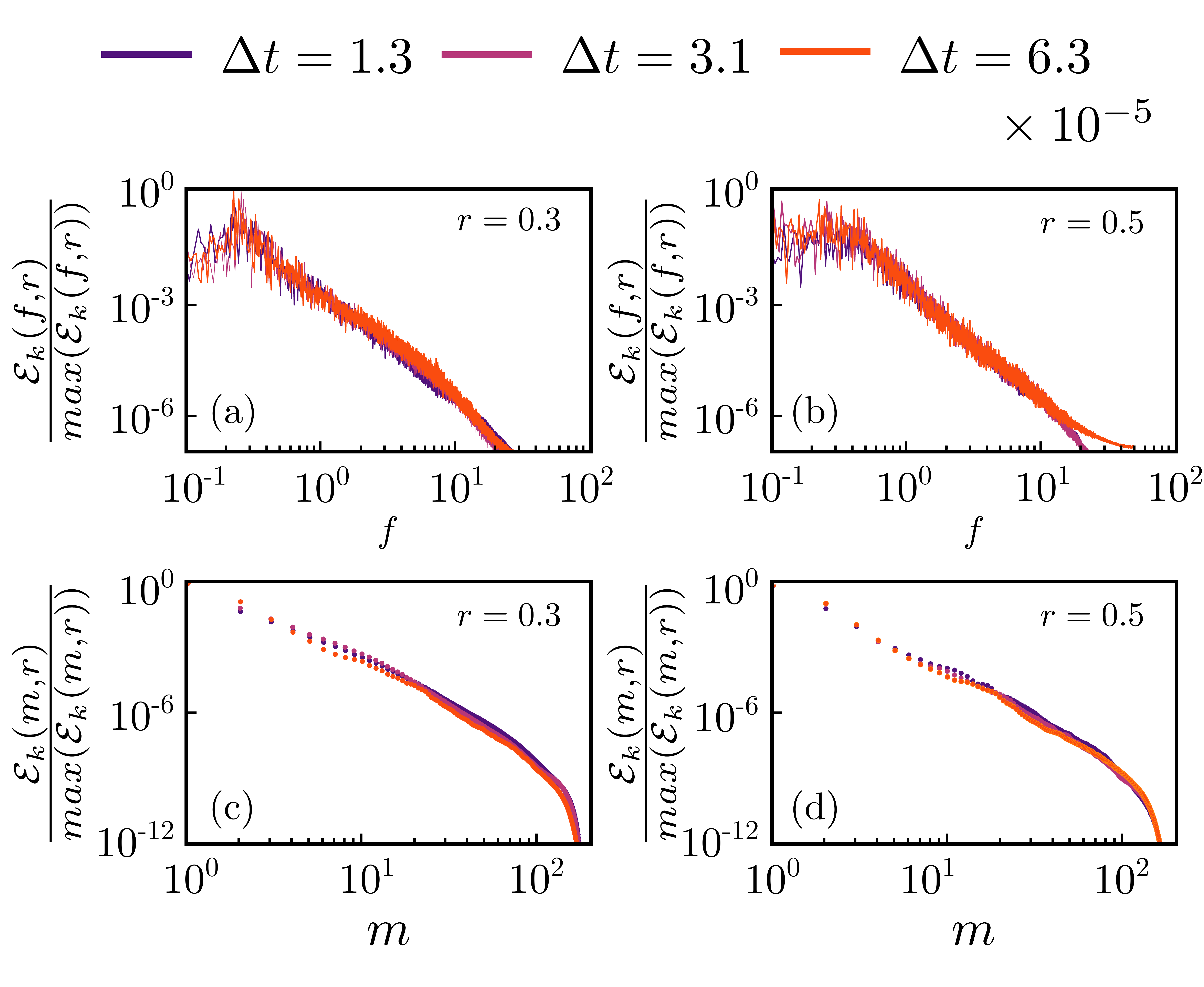}
    \caption{Time step independence study testing three time steps $\Delta t$ for the fine mesh at $Wi=25$. See Fig.\ \ref{fig:GCS} for a description of the panels.}
    \label{fig:TCS}
\end{figure}

In Fig.~\ref{fig:GCS}, we show both the temporal and spatial kinetic energy spectra for the four meshes at $Wi=25$ and two radial locations ($r=0.3$ and $r=0.5$). From the temporal spectra at $r=0.5$, only the coarse mesh shows spikes at high frequencies (analogous to those observed in Ref.~\cite{buel_elastic_2018}). The power-law decay exponent $\zeta_k$ within $r_{BL,k}$ also shows a consistent trend for the fine and extra fine meshes, except for minor quantitative differences that depend on the selection of the power-law fitting range.

The temporal convergence of our simulations is further assessed by varying the time step. Figure~\ref{fig:TCS} reports the kinetic energy spectra computed for $\Delta t \in \left( 1.3,\ 3.1,\ 6.3 \right) \times  10^{-5}$ using fine mesh. The results show that $\Delta t =6.3\times10^{-5}$ is sufficiently small not to induce numerical artifacts.

\subsection*{Appendix B: Elastic and kinetic energy spectra}
\begin{figure}[t]
    \centering
       \includegraphics[width=0.675\linewidth]{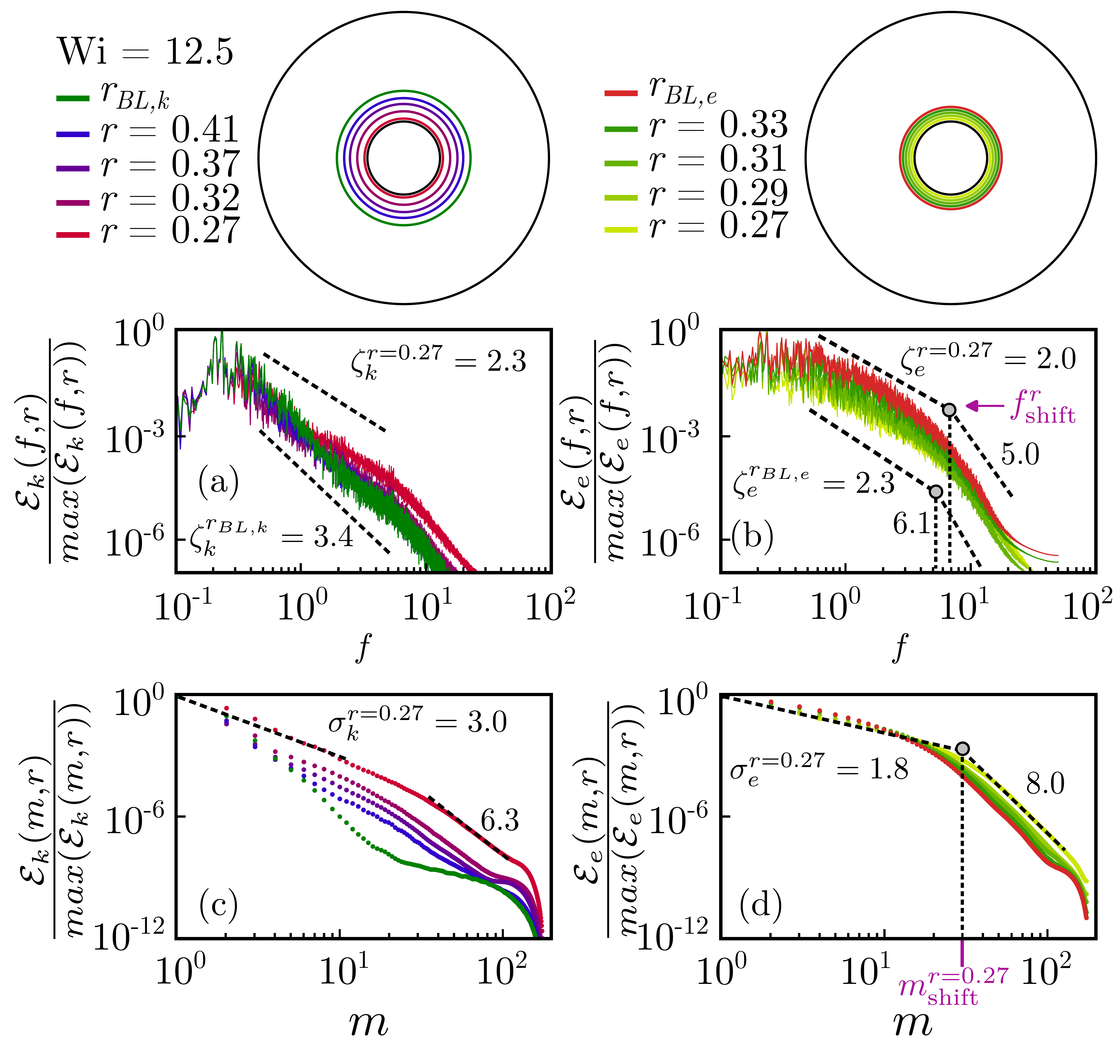}
    \caption{Temporal (a,b) and spatial (c,d) spectra of kinetic (a,c) and elastic (b,d) energy at different radial positions inside their respective boundary layers for $Wi=12.5$. The same color coding as in Fig.~\ref{fig:spectra_Wi25} is used.}
    \label{fig:spectra_Wi12}
\end{figure}

\begin{figure}[t]
    \centering
       \includegraphics[width=0.675\linewidth]{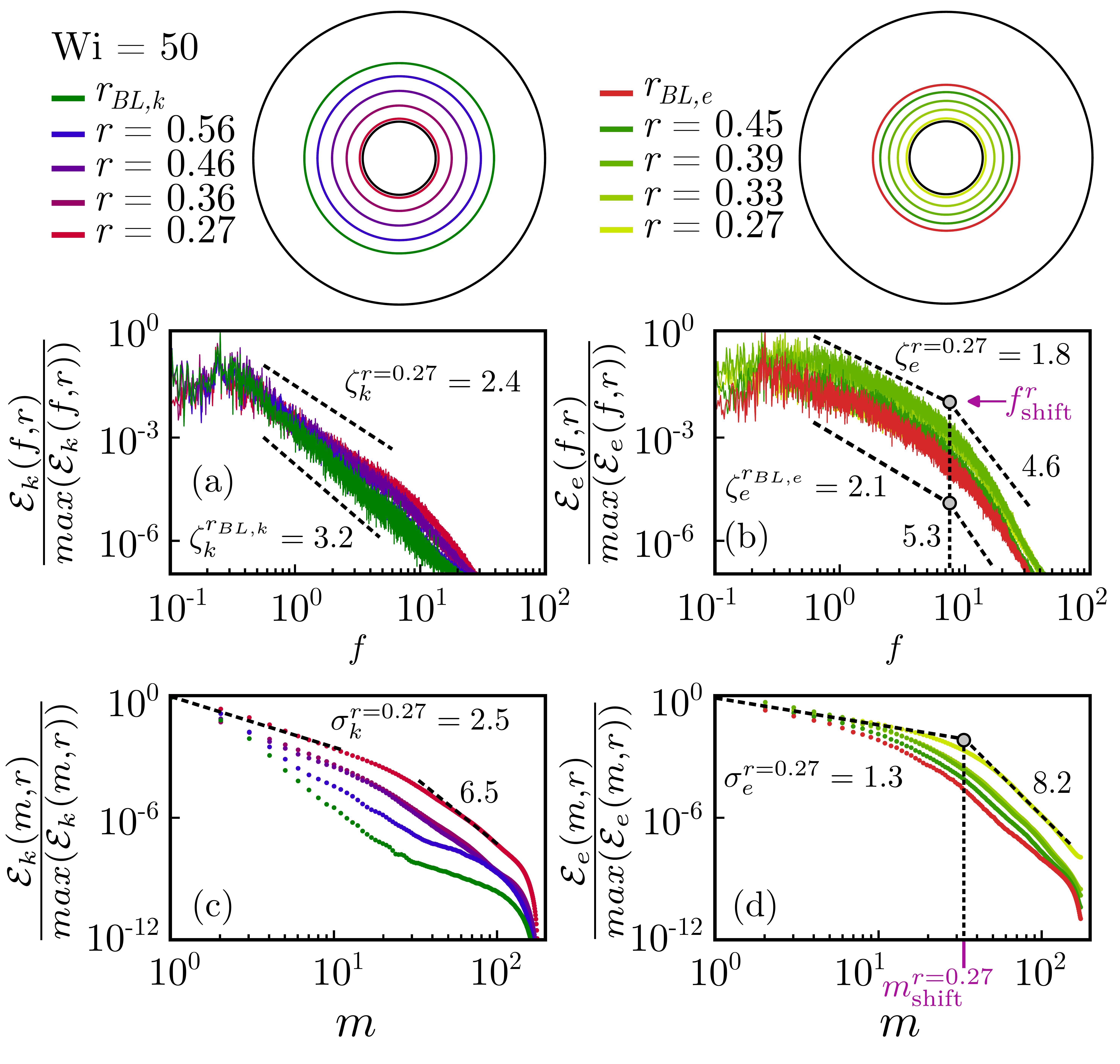}
    \caption{Temporal (a,b) and spatial (c,d) spectra of kinetic (a,c) and elastic (b,d) energy at different radial positions inside their respective boundary layers for $Wi=50$. The same color coding as in Fig.~\ref{fig:spectra_Wi25} is used.}
    \label{fig:spectra_Wi50}
\end{figure}

\begin{figure}[t]
    \centering
    \centering
       \includegraphics[width=0.675\linewidth]{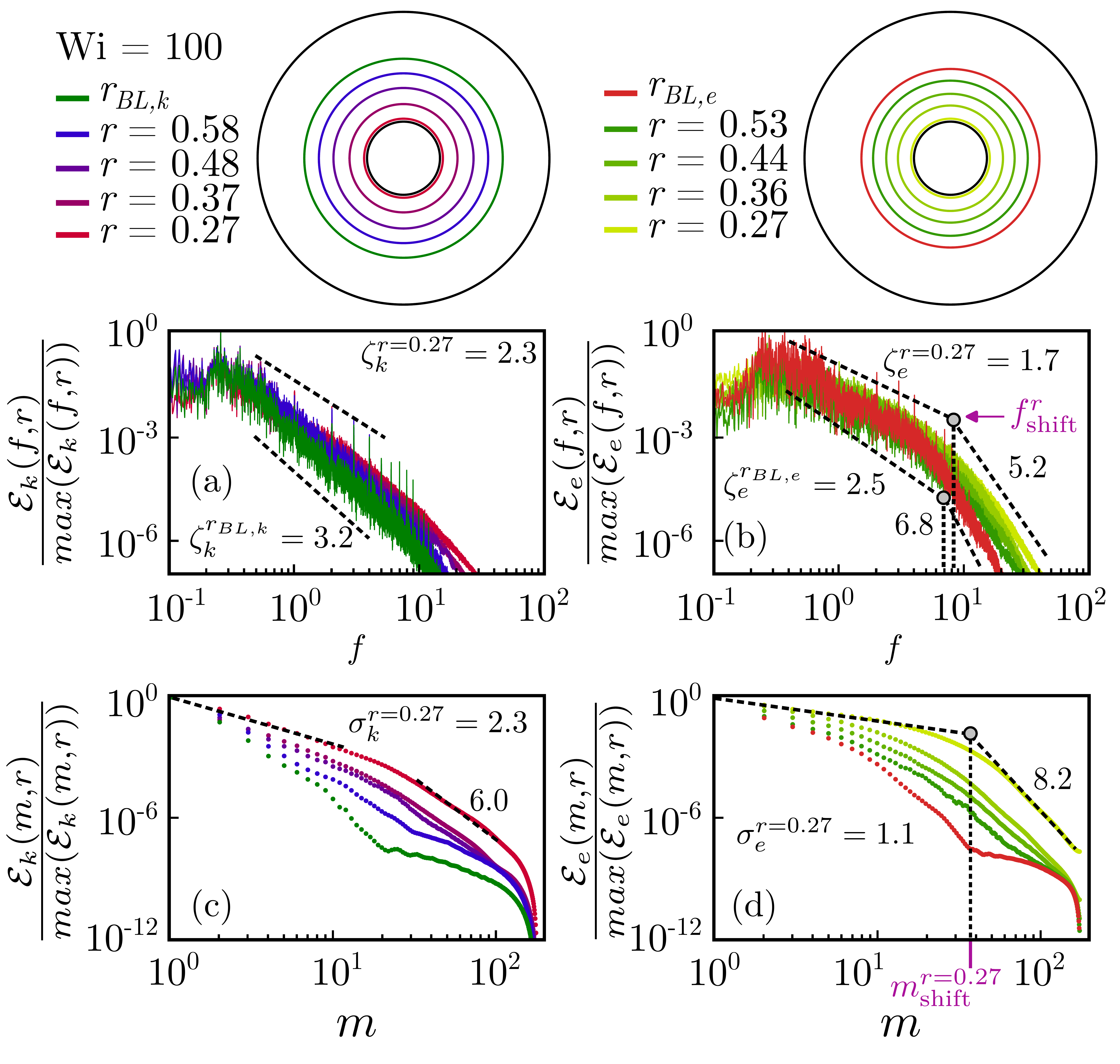}
    \caption{Temporal (a,b) and spatial (c,d) spectra of kinetic (a,c) and elastic (b,d) energy at different radial positions inside their respective boundary layers for $Wi=100$. The same color coding as in Fig.~\ref{fig:spectra_Wi25} is used.}
    \label{fig:spectra_Wi100}
\end{figure}

The elastic and kinetic energy temporal and spatial spectra (computed with the fine grid) are shown in Figs.~\ref{fig:spectra_Wi12}, \ref{fig:spectra_Wi50}, and \ref{fig:spectra_Wi100} for $Wi = 12.5$, 50 and 100, respectively. The radial positions at which they are computed are illustrated in the top lines of the same figures. The picture emerging from these plots confirms that of Sec.~\ref{sec:spectra} for $Wi=25$, with only minor differences at the highest Weissenberg number.


 \ \\ \ 

\color{black}

\end{document}